\documentclass[apjl]{emulateapj}
\usepackage{psfig,amsfonts,amsmath,graphicx,natbib,apjfonts,lscape,nicefrac}

\newcommand{\mum}{$\rm{\mu}$m}
\newcommand{\spitzer}{\textit{Spitzer}}

\shorttitle{Mass-Metallicity Relation at $z\sim 3-5$}
\shortauthors{Laskar et al.}

\def\cfa{1}
\def\ssc{2}

\begin{document} 

\title{Exploring the Galaxy Mass-Metallicity Relation at $z\sim 3-5$}

\author{
Tanmoy Laskar\altaffilmark{\cfa},
Edo Berger\altaffilmark{\cfa},
and Ranga-Ram Chary\altaffilmark{\ssc}
}

\altaffiltext{\cfa}{Harvard-Smithsonian Center for Astrophysics, 60
Garden Street, Cambridge, MA 02138}

\altaffiltext{\ssc}{Spitzer Science Center, California Institute of
Technology, Pasadena, CA 91125}

\begin{abstract}
Long-duration gamma-ray bursts (GRBs) provide a premier tool for
studying high-redshift star-forming galaxies thanks to their extreme
brightness and association with massive stars.  Here we use GRBs to
study the galaxy stellar mass-metallicity ($M_*$-$Z$) relation at $z\sim 3-5$,
where conventional direct metallicity measurements are extremely
challenging. We use the interstellar medium metallicities of long GRB
hosts derived from afterglow absorption spectroscopy ($Z\approx
0.01-1$ Z$_\odot$), in conjunction with host galaxy stellar masses
determined from deep \spitzer\ 3.6 $\mu$m observations of 20 GRB
hosts.  We detect about 1/4 of the hosts with $M_{\rm AB}(I)\approx
-21.5$ to $-22.5$ mag, and place a limit of $M_{\rm AB}(I)\gtrsim
-19$ mag on the remaining hosts from a stacking analysis.  Using these
observations, we present the first rest-frame optical luminosity
distribution of long GRB hosts at $z\gtrsim 3$ and find that it is
similar to the distribution of long GRB hosts at $z\sim 1$.  In
comparison to Lyman-break galaxies at the same redshift, GRB hosts are
generally fainter, but the sample is too small to rule out an overall
similar luminosity function.  On the other hand, the GRB hosts appear
to be more luminous than the population of Lyman-alpha emitters at
$z\sim 3-4$.  Using a conservative range of mass-to-light ratios for
simple stellar populations (with ages of 70 Myr to $\sim 2$ Gyr), we
infer the host stellar masses and present mass-metallicity
measurements at $z\sim 3-5$ ($\langle z\rangle \approx 3.5$).  We find
that the detected GRB hosts, with $M_*\approx 2\times 10^{10}$
M$_\odot$, display a wide range of metallicities, but that the mean
metallicity at this mass scale, $Z\approx 0.1$ Z$_\odot$, is lower
than measurements at $z\lesssim 3$.  Combined with stacking of the
non-detected hosts with $M_*\lesssim 3\times 10^9$ M$_\odot$ and
$Z\lesssim 0.03$ Z$_\odot$, we find evidence for the existence of an
$M_*$-$Z$ relation at $z\sim 3.5$ and continued evolution of this
relation to systematically lower metallicities from $z\sim 2$.
\end{abstract}

\section{Introduction}

The simple ``closed-box'' model of galaxy evolution \citep{Talbot1971}
predicts a correlation between the stellar mass and the gas-phase
metallicity of a galaxy (the $M_*$-$Z$ relation), under the assumptions
of no gas inflows or outflows, a constant yield of metals, an
invariant stellar initial mass function (IMF), and instantaneous
mixing of newly-synthesized metals back into the interstellar medium
(ISM).  In reality, this simple picture is complicated by the fact
that galaxies accrete low-metallicity gas from the intergalactic
medium (IGM) and lose metal-enriched gas through galactic-scale winds
or by depletion on to dust.  In addition to these processes, the
$M_*$-$Z$ relation may also be modified by a mass-dependent
star-formation efficiency 
\citep{Juneau2005,Feulner2005,Franceschini2006,Asari2007}, and
possibly an environmental-dependent IMF \citep{Koppen2007}.  Thus, the
$M_*$-$Z$ relation and its evolution with redshift provide insight into
the physical processes that shape galaxy formation and evolution
across cosmic time.

Given the importance of this relation it has been the focus of several
extensive studies out to $z\sim 3$.  In the local universe ($z\sim
0.1$), \citet{Tremonti2004} studied 53,400 galaxies from the Sloan
Digital Sky Survey and found a tight correlation ($\pm 0.1$ dex)
between stellar mass and metallicity over a range of $M_*\approx
10^{8.5-11.5}$ M$_\odot$ and an order of magnitude in metallicity (see
also \citealt{Kewley2008}).  They concluded that the observed
correlation is best explained by the influence of metal-enriched
outflows, with a larger metal loss in lower mass galaxies.  Studies at
$z\sim 1-3$
\citep{Savaglio2005,Erb2006,Maiolino2008,Mannucci2009,jkb10} found that
the $M_*$-$Z$ relation evolves by about $0.8$ dex from $z\sim 3$ to the
present, while keeping the same overall trend.  \citet{Savaglio2005}
argued that the redshift evolution to $z\sim 0.7$ can be reproduced in
the simple closed-box model with the assumption that the typical
timescale for star formation is longer in lower mass galaxies.  On the
other hand, \citet{jkb10} found that the $M_*$-$Z$ relation evolves only
below $\sim 10^{10.5}$ M$_\odot$ to $z\sim 0.8$, and argued that unlike
in the local universe the effective yield decreases with larger mass
and that a closed-box model cannot explain the evolution.  They
further argued that outflows play a minor role, and proposed that a
rising star formation efficiency with large mass may be the dominant
effect.  \citet{Erb2006} argued that at $z\sim 2.3$ the primary driving
mechanism for the $M_*$-$Z$ relation and its evolution is the expected
increase in metallicity as star formation leads to a reduced gas
fraction, and that outflows affect galaxies at all mass scales.

Beyond $z\sim 2.3$ there are only a few measurements of galaxy
metallicities and masses.  \citet{Mannucci2009} studied 10 Lyman break
galaxies (LBGs) at $z\sim 3.1$ and found continued downward evolution
of the $M_*$-$Z$ relation, and a decreasing effective yield with larger
stellar mass.  They argued that gas infall plays the dominant role in
the $M_*$-$Z$ relation, and that outflows are not needed.
\citet{Maiolino2008} studied 9 LBGs at $z\sim 3.5$ and found a decline
in the mean metallicity at a stellar mass scale of $\sim 1.4\times
10^{10}$ M$_\odot$ compared to $z\sim 2.3$, with a possible steepening
of the $M_*$-$Z$ relation relative to lower redshifts.

Tracing the $M_*$-$Z$ relation and its evolution to even earlier times
will provide insight into the earliest epochs of galaxy evolution,
while allowing us to probe the relative importance of the various
galactic-scale phenomena proposed at $z\lesssim 3$.  Although initial
studies of LBGs at $z\sim 3.5$ are now available
\citep{Maiolino2008,Mannucci2009}, these studies are challenging
because the nebular emission lines required for robust metallicity
measurements\footnotemark\footnotetext{For example,
\citet{Mannucci2009} use the $R_{23}$ diagnostic to determine
metallicities for their $z\sim 3$ galaxy sample, but this relation is
known to be double-valued.  They attempt to discriminate between the
low- and high-metallicity branches using the \ion{O}{3}$\lambda
5007$/\ion{O}{2}$\lambda 3727$ ratio.}  (e.g., H$\alpha$, H$\beta$,
\ion{N}{2}$\lambda 6583$, \ion{O}{3}$\lambda\lambda 4959,5007$,
\ion{O}{2}$\lambda\lambda 3726,3729$) shift into the near- and mid-IR,
where existing spectrographs have reduced sensitivity compared to the
optical band.  This is further complicated by the rapid dimming of
galaxies at higher redshift such that only the most luminous LBGs are
amenable to spectroscopy.

An alternative way to determine metallicities at $z\gtrsim 3$ (and in
principle at $z\sim 10$ and beyond; \citealt{Salvaterra2009,Tanvir2009}) is
absorption spectroscopy of gamma-ray burst (GRB) optical/near-IR
afterglows.  Long-duration GRBs are known to be associated with the
deaths of massive stars (e.g., \citealt{Woosley2006}), and therefore
with sites of active star formation.  The large optical luminosities
of GRB afterglows (easily exceeding 20 mag for several hours even at
$z\sim 8$; \citealt{Tanvir2009}), and their intrinsic featureless spectra,
provide a unique way to measure interstellar medium (ISM)
metallicities for galaxies at $z\gtrsim 2$ from rest-frame ultraviolet
metal absorption lines and Ly$\alpha$ absorption.  Since the afterglows are
significantly brighter than the underlying host galaxies, this
technique allows us to measure metallicities independent of the galaxy
brightness.  Moreover, since long GRB progenitors reside in star
forming environment within their hosts, their sight-lines probe the
warm ISM and \ion{H}{2} regions that give rise to the (rest-frame
optical) nebular emission lines that are used for metallicity
measurements at $z\lesssim 3$.  This approach has now been exploited
at least to $z\sim 5$ using optical spectra (e.g.,
\citealt{Berger2006,Prochaska2007,Fynbo2009}), and with near-IR spectrographs it
can be implemented to $z\sim 20$.

Naturally, to explore the $M_*$-$Z$ relation at $z\gtrsim 3$ we also
require a determination of the GRB host galaxy stellar masses, and
hence follow-up infrared observations with the \spitzer\ Space
Telescope to probe the rest-frame optical luminosity.  Here, we
present the first large set of \spitzer\ observations for GRB host
galaxies at $z\sim 3-5$, and combine the inferred masses with measured
metallicities to explore the $M_*$-$Z$ relation beyond $z\sim 3$.  Since
deep \spitzer/IRAC images are generally confusion-limited, our use of
GRB afterglows provides an additional boon --- they accurately
pinpoint the location of the host galaxies (to $\sim 0''.1$), thereby
allowing for accurate galaxy
identifications\footnotemark\footnotetext{This can be contrasted with
the potential use of quasar {\it intervening} absorption systems for
studies of the $M_*$-$Z$ relation, since the galaxy counterparts of the
absorbers are offset on the sky by $\sim {\rm few}$ arcseconds from
the quasar position.  Spectroscopic confirmation is therefore required
to determine the correct counterpart, negating the advantage of
absorption spectroscopy.  Furthermore, even if the galaxy counterparts
could be identified, observations with {\it Spitzer}'s large
point-spread-function against the much brighter quasar glare are
essentially impossible.}.  The plan of the paper is as follows.  We
present the \spitzer\ observations, analysis, photometry, 
and metallicity data in \S\ref{sec:sample}.  In \S\ref{sec:lum} we
present the first rest-frame optical luminosity distribution of GRB
hosts at $z\gtrsim 3$ and compare it to both $z\sim 1$ GRB hosts and
field galaxy samples at $z\sim 3$.  We derive the mass distribution in
\S\ref{sec:mass}.  Finally, in \S\ref{sec:mz} we combine the mass and
metallicity measurements to place the first points on the $M_*$-$Z$
diagram at $z\gtrsim 3$.  We explore the implications of our results
and future prospects in \S\ref{sec:conc}.

\section{GRB Sample and Data Analysis} 
\label{sec:sample}

We obtained deep observations of all 35 long
GRB host galaxies with spectroscopic redshifts in the range $z\approx
2-5.8$ available as of November 2006 using the 3.6 \mum\ band of the
Infra-Red Array Camera (IRAC; \citealt{Fazio2004}) on-board the {\it
Spitzer} Space Telescope.  Here we investigate the properties of GRB
hosts in the redshift range $z\approx 3-5.6$ (observed for about 2 
hours each); targets at redshifts $z\approx 2-3$ are treated elsewhere
(Chary et al.~2011, in prep).  For the objects in this paper, the effective
wavelength of the IRAC 3.6 \mum\ band probes the rest-frame spectral
energy distribution (SED) redward of about 5500 \AA\
(Figure~\ref{fig:z}) and therefore provides a robust measure of the
stellar mass.

We processed the \spitzer\ data using the standard \textsc{mopex}
\citep{Makovoz2006} software package to generate mosaics for each
target.  The \textsc{mopex} package detects and removes cosmic rays
and moving objects before drizzling \citep{Fruchter2002}, performing
background equalization, and applying distortion corrections.  For our
coverage and dither pattern we find that an output pixel scale of
$0''.4$ and a drizzling parameter of 0.7 provide the best combination
of improvement in the point-spread-function (PSF) with minimal
degradation of the signal-to-noise ratio.  We set all other parameters
in \textsc{mopex} to their recommended defaults.

\subsection{Astrometry}

\begin{figure}[ht]
\begin{center}
\begin{minipage}{0.24\columnwidth}
\includegraphics[width=\columnwidth,angle=-90]{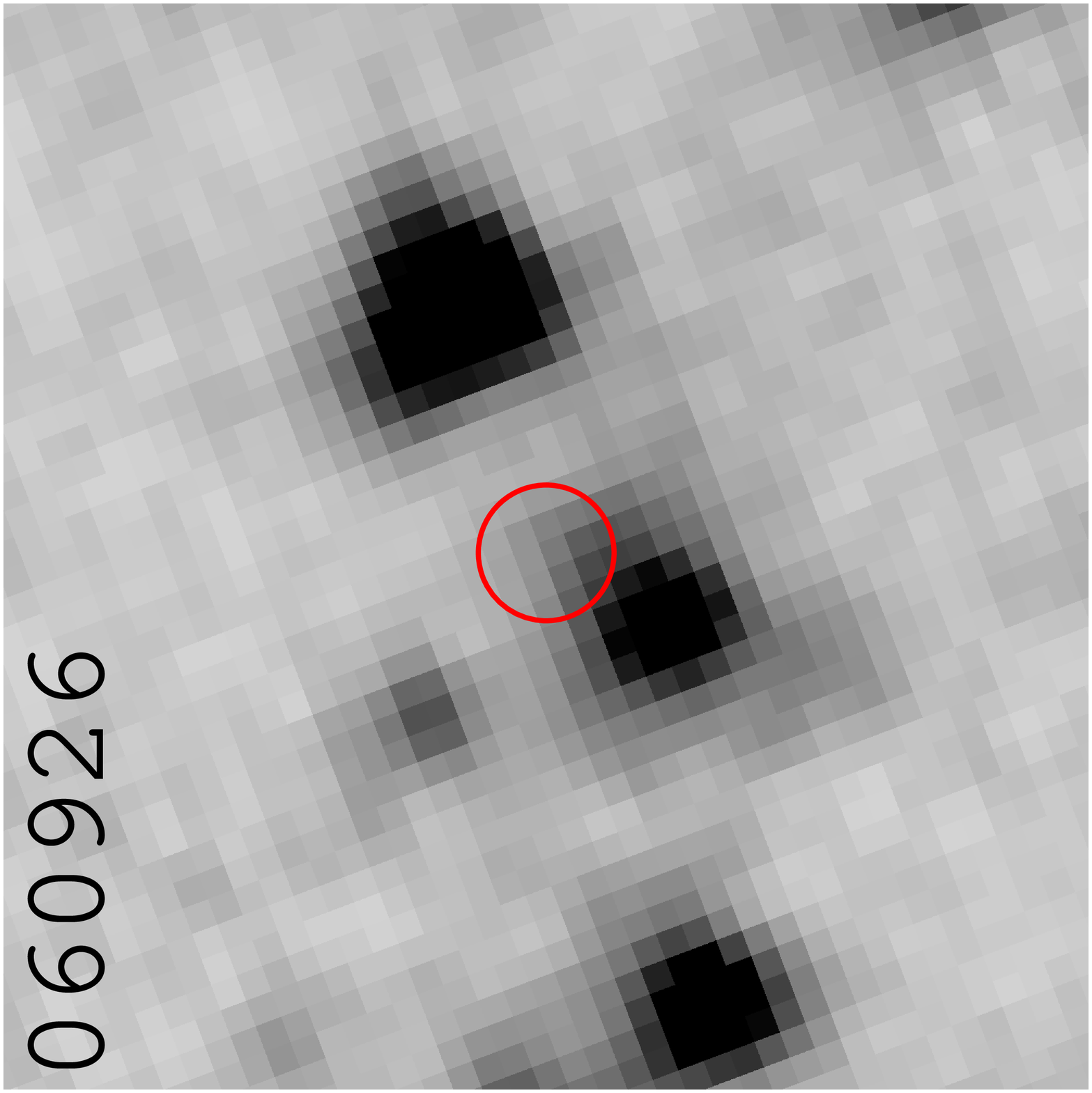}
\end{minipage}
\begin{minipage}{0.24\columnwidth}
\includegraphics[width=\columnwidth,angle=-90]{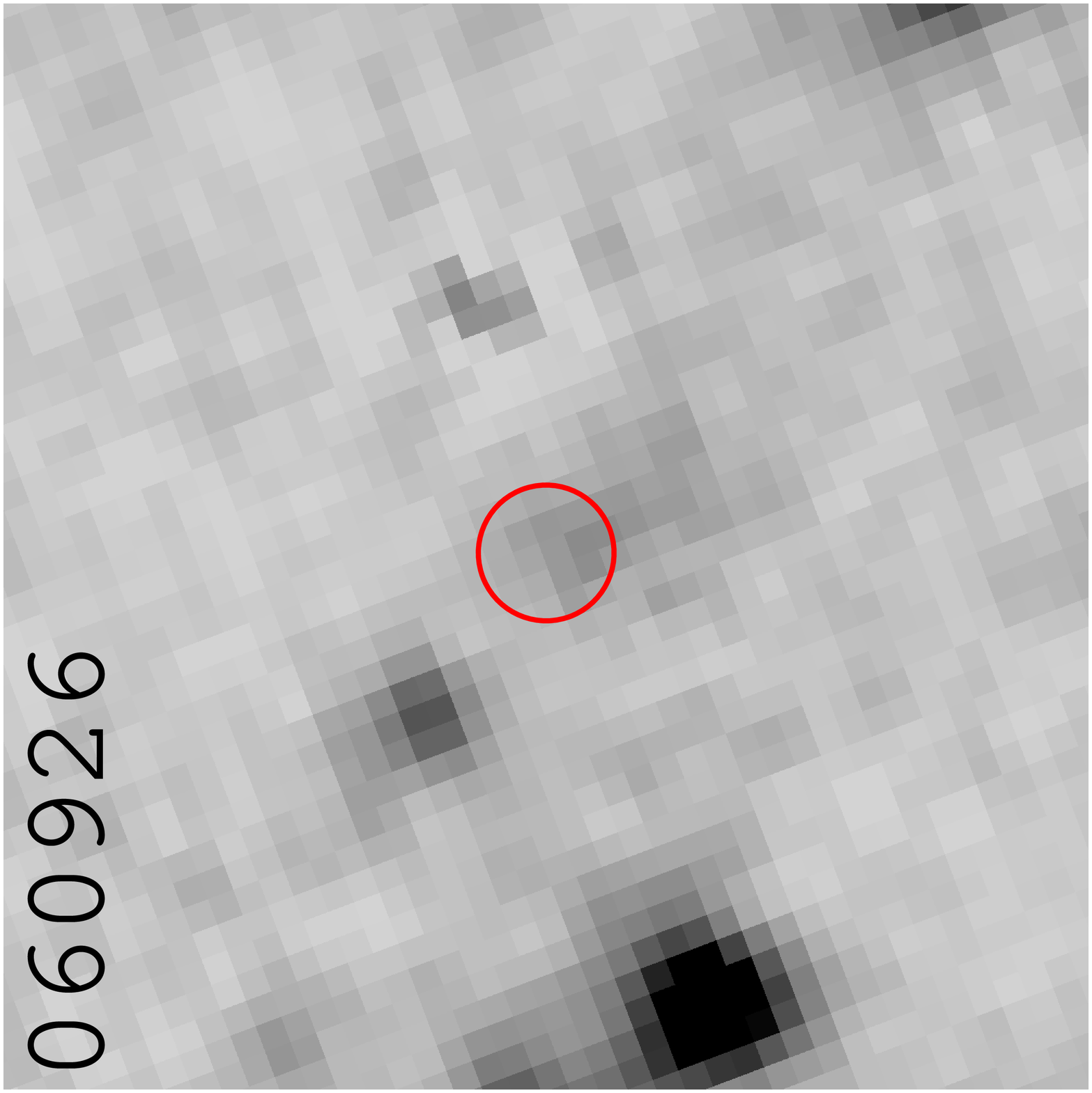}
\end{minipage}\\
\end{center}
\begin{center}
\begin{minipage}{0.24\columnwidth}
\includegraphics[width=\columnwidth,angle=-90]{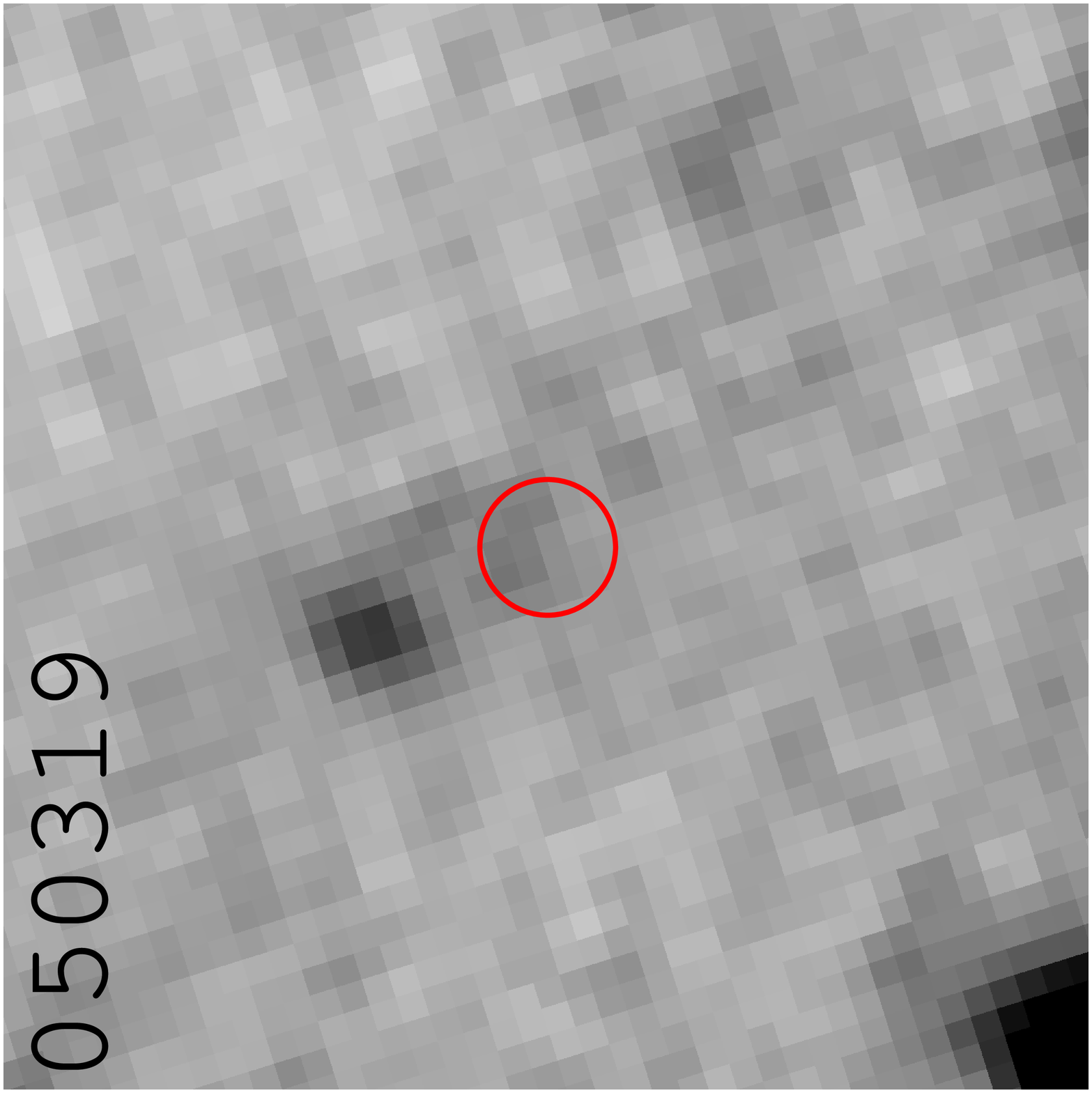}
\end{minipage}
\begin{minipage}{0.24\columnwidth}
\includegraphics[width=\columnwidth,angle=-90]{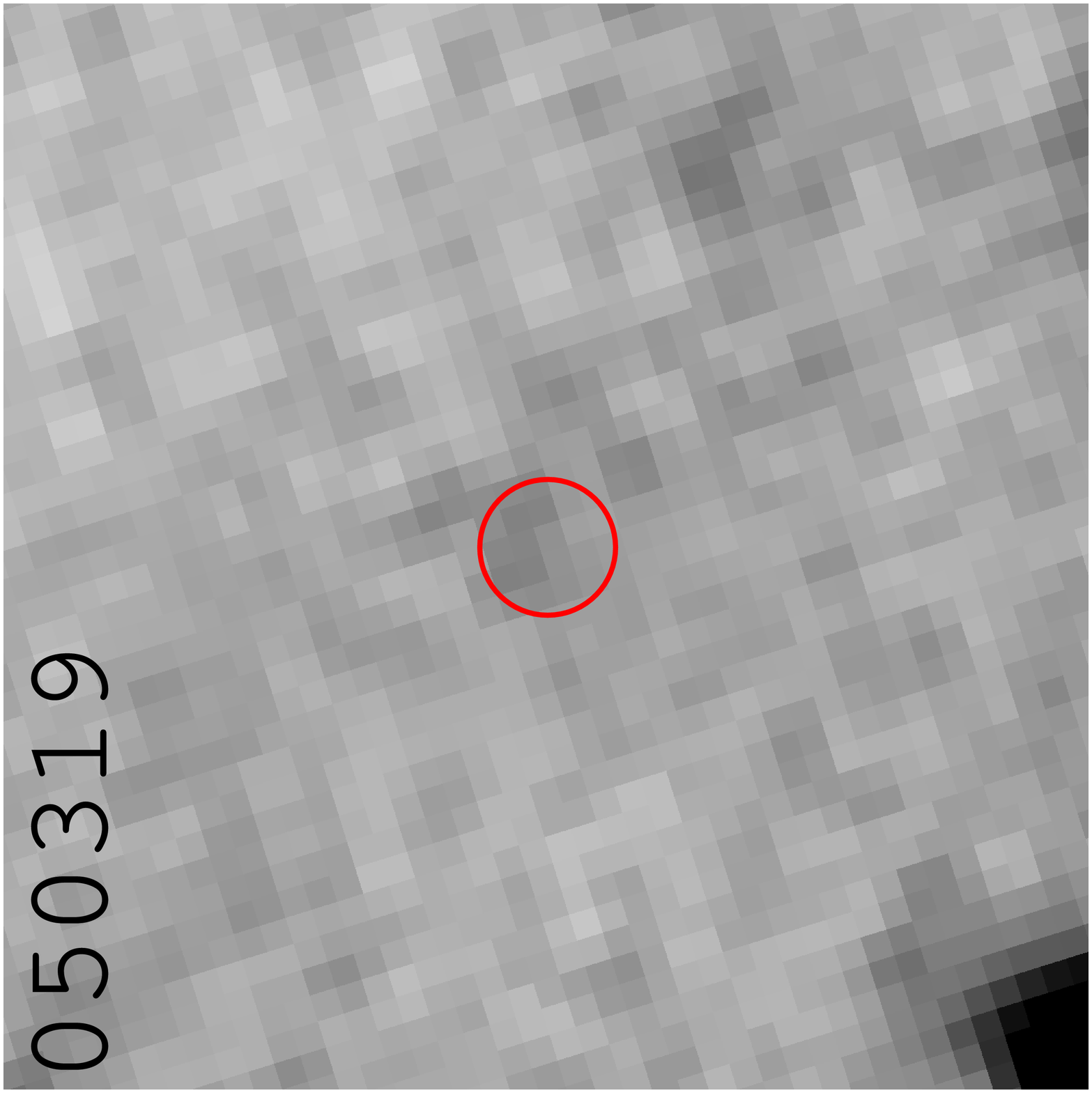}
\end{minipage}\\
\end{center}
\begin{minipage}{0.24\columnwidth}
\hspace{1.813in}\includegraphics[width=\columnwidth,angle=-90]{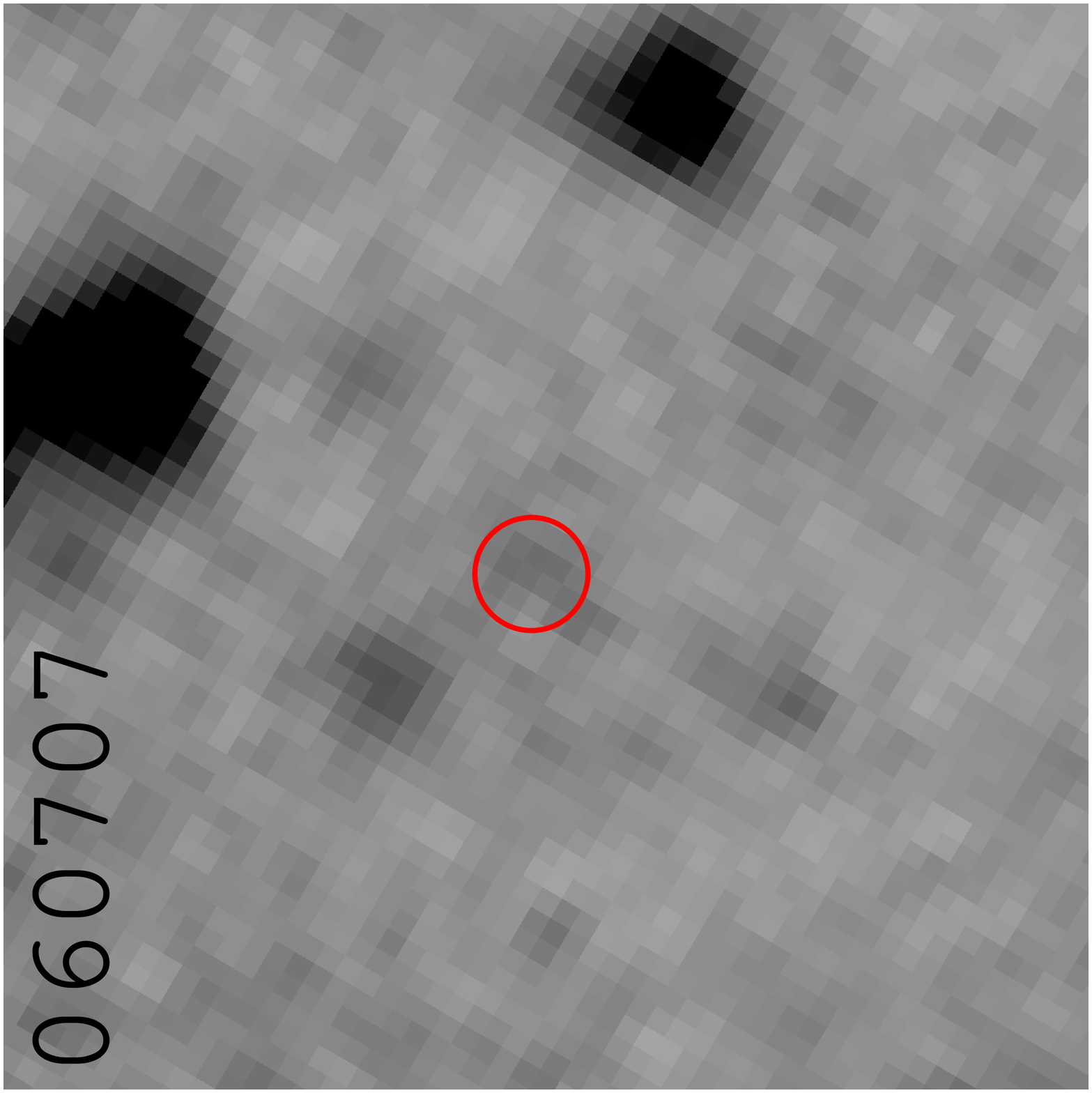}
\end{minipage}\\
\begin{center}
\begin{minipage}{0.24\columnwidth}
\includegraphics[width=\columnwidth,angle=-90]{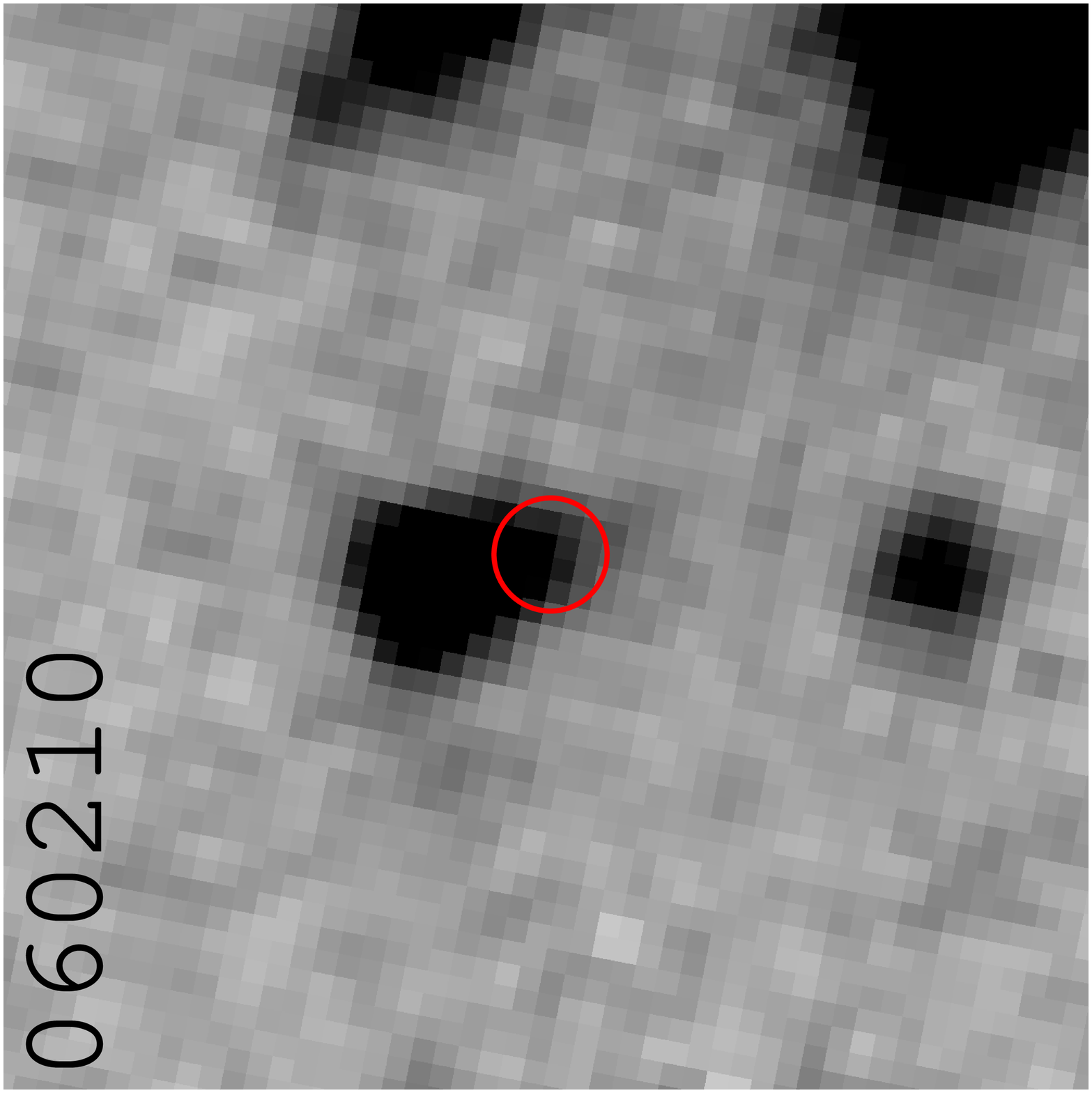}
\end{minipage}
\begin{minipage}{0.24\columnwidth}
\includegraphics[width=\columnwidth,angle=-90]{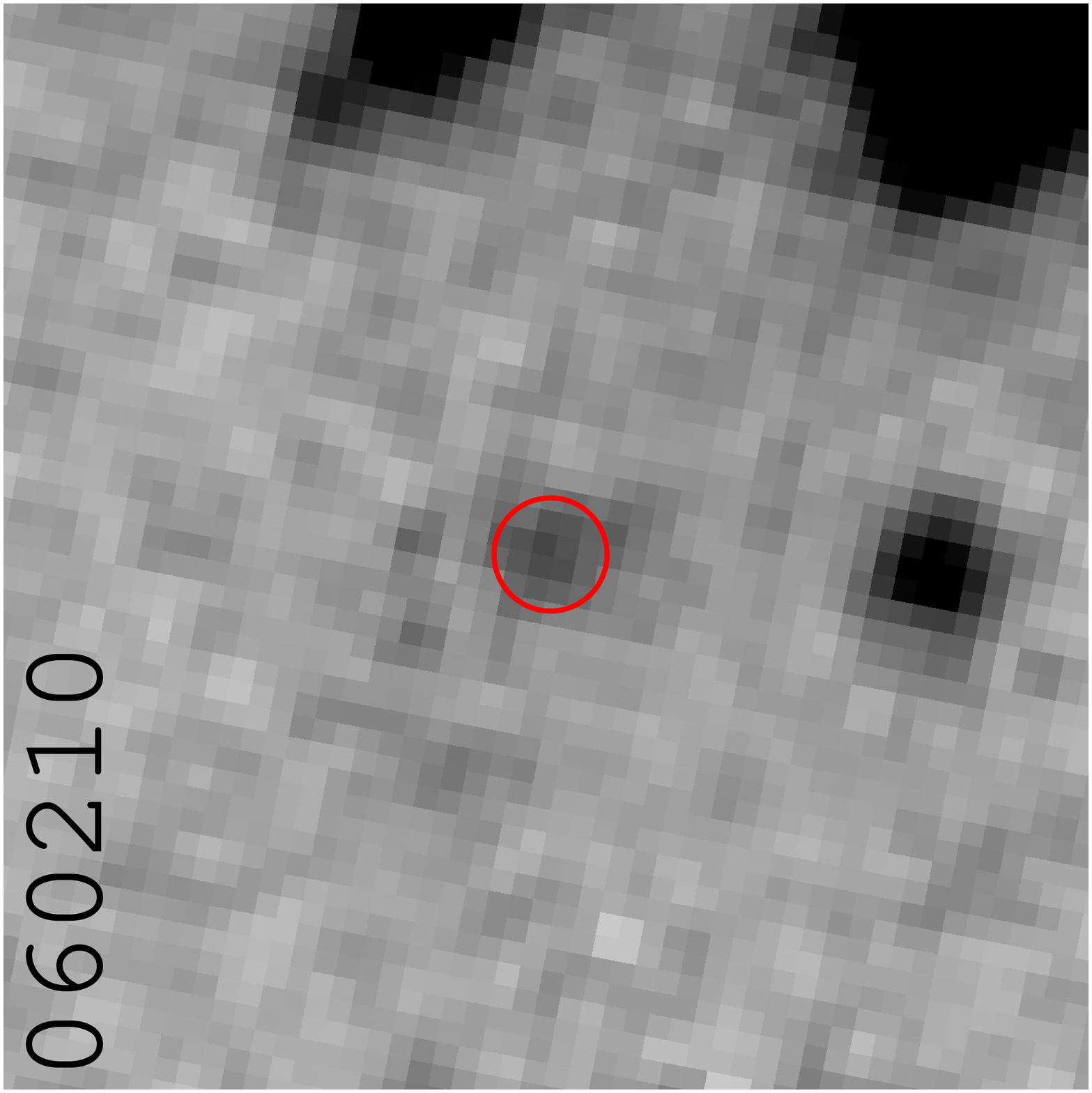}
\end{minipage}\\
\end{center}
\begin{minipage}{0.24\columnwidth}
\hspace{1.813in}\includegraphics[width=\columnwidth,angle=-90]{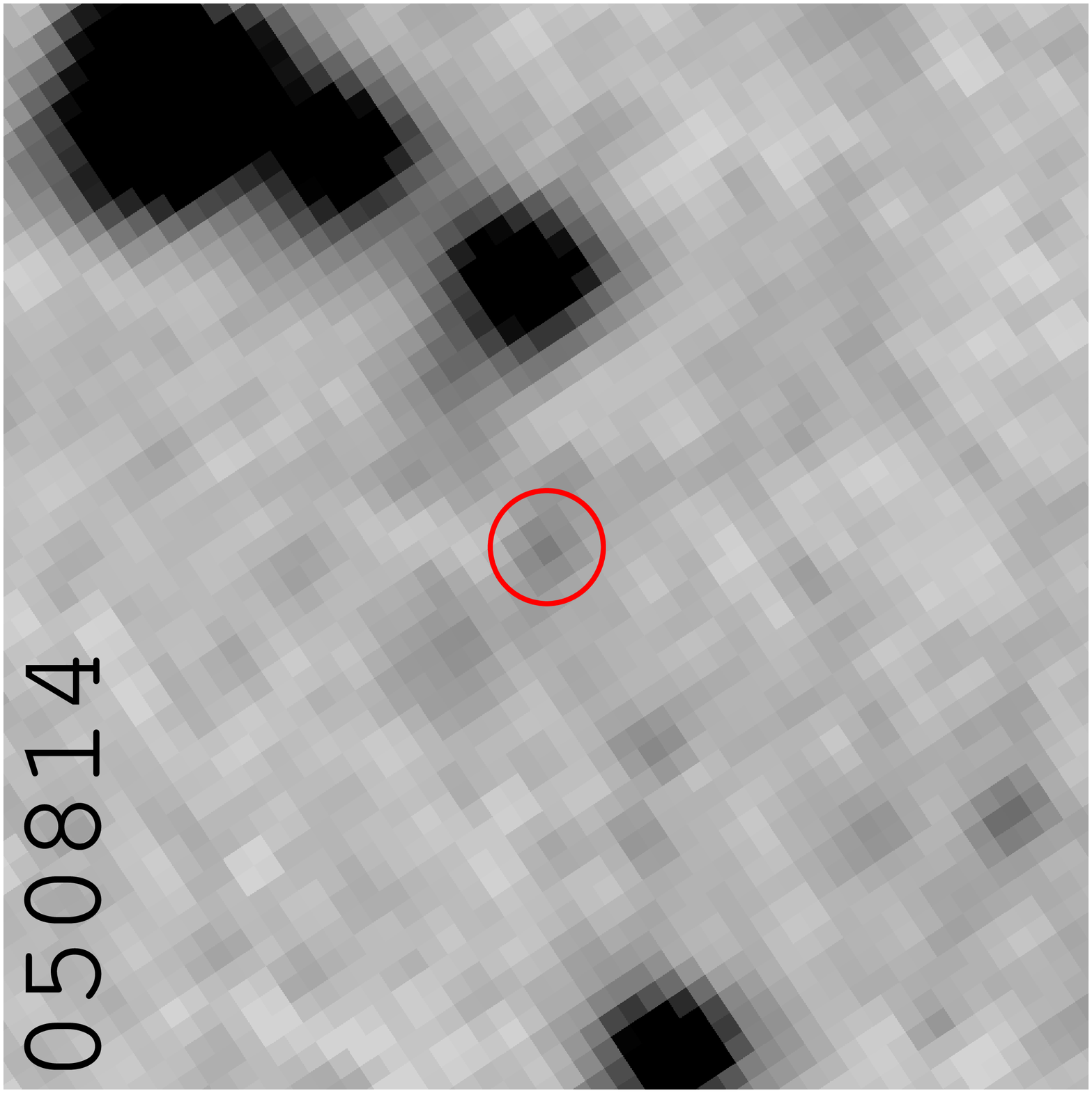}
\end{minipage}
\caption{\spitzer\ images of the GRB host galaxies with 3.6 \mum\
detections.  The left-hand panels show the processed images, while the
right-hand panels include subtractions of nearby sources using
\textsc{galfit} (when performed).  The circles (1\arcsec\ radius) mark
the afterglow positions.  All images have the same orientation (North
is up and East is to the left) and scale (16\arcsec\ on a side) with
0\arcsec.4 square pixels.  
\label{fig:cutouts1}}
\end{figure}

\begin{figure}[ht]
\begin{center}
\includegraphics[width=0.24\columnwidth,angle=-90]{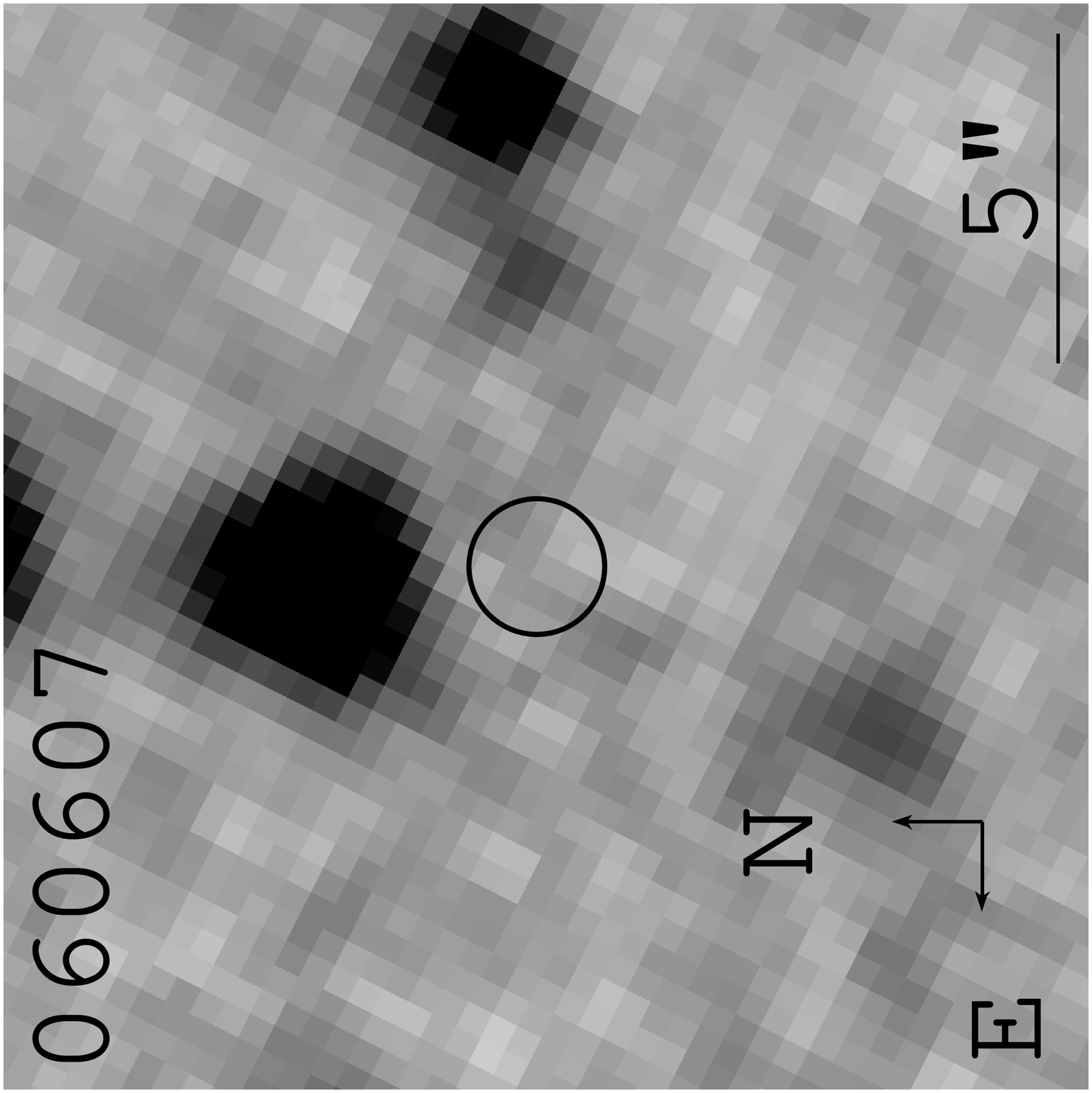}
\includegraphics[width=0.24\columnwidth,angle=-90]{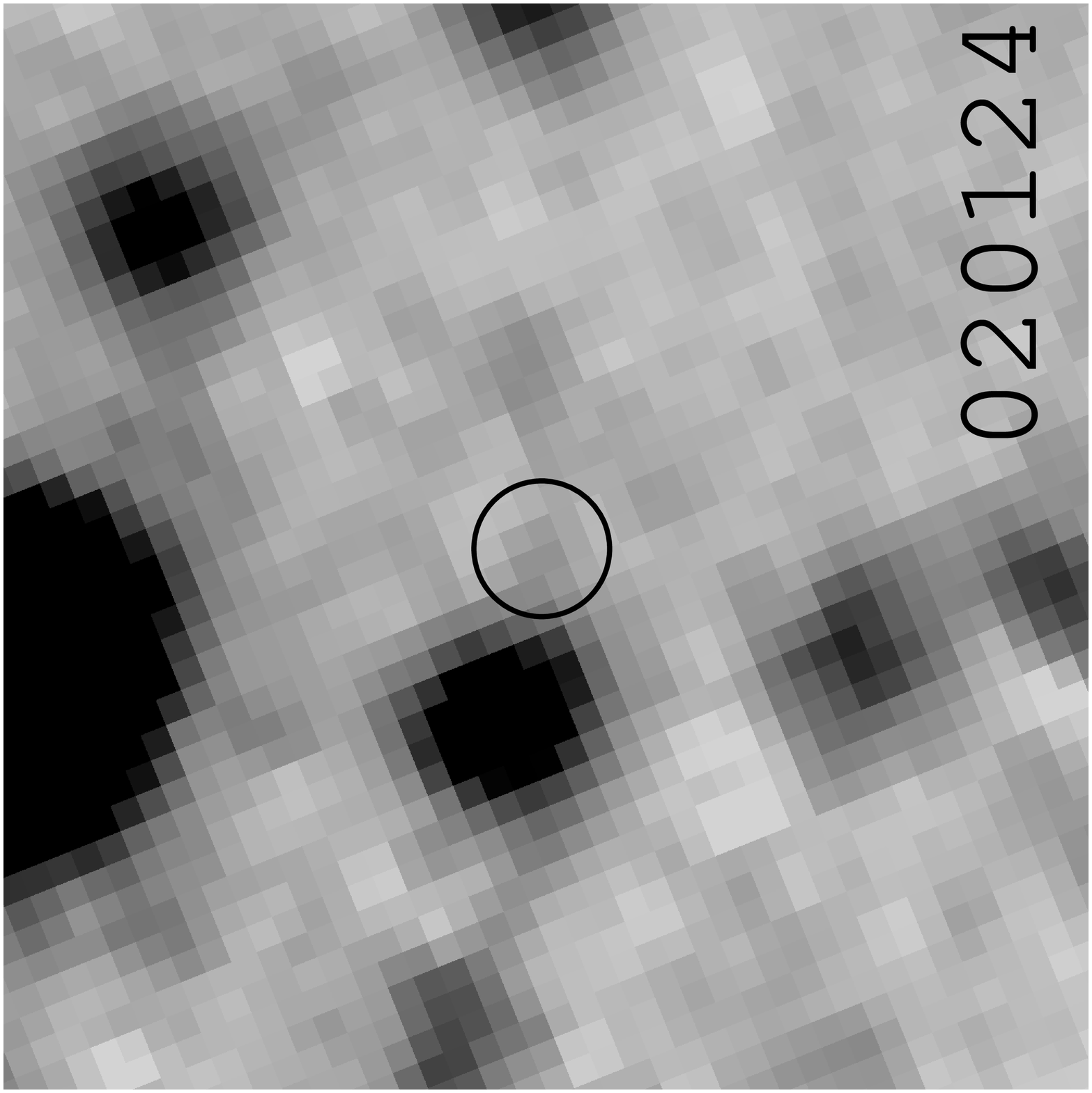}
\includegraphics[width=0.24\columnwidth,angle=-90]{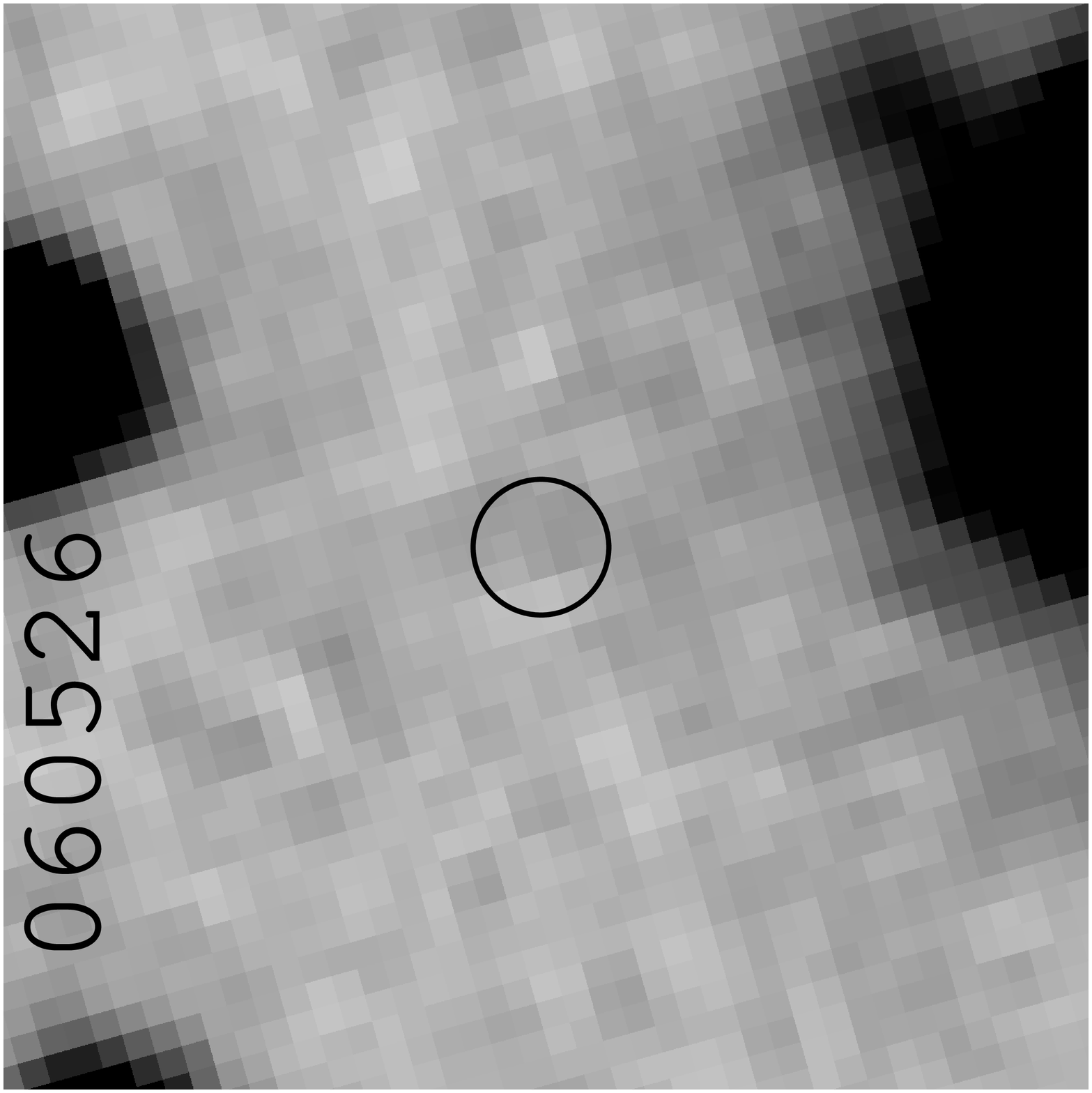}\\
\includegraphics[width=0.24\columnwidth,angle=-90]{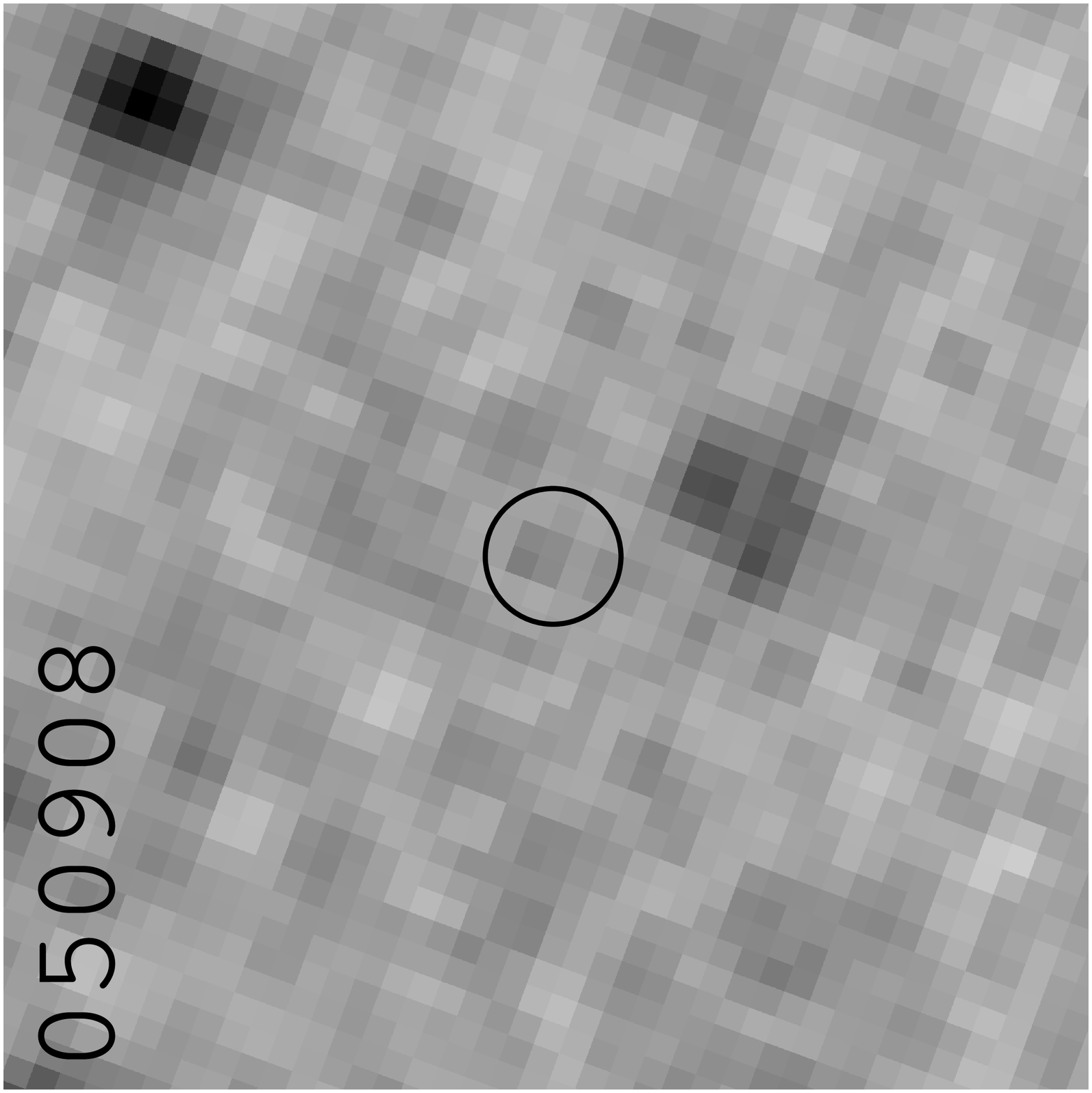}
\includegraphics[width=0.24\columnwidth,angle=-90]{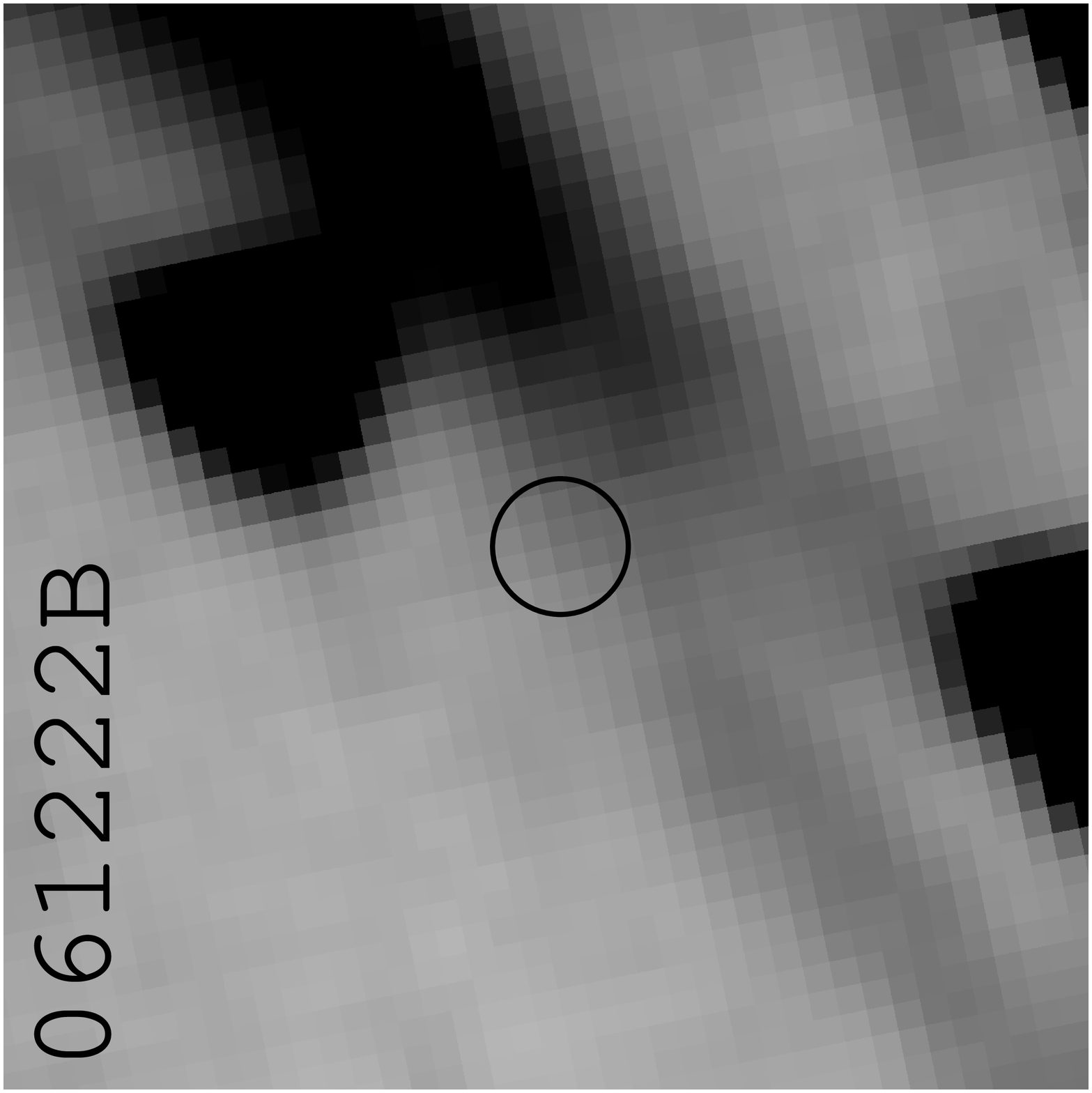}
\includegraphics[width=0.24\columnwidth,angle=-90]{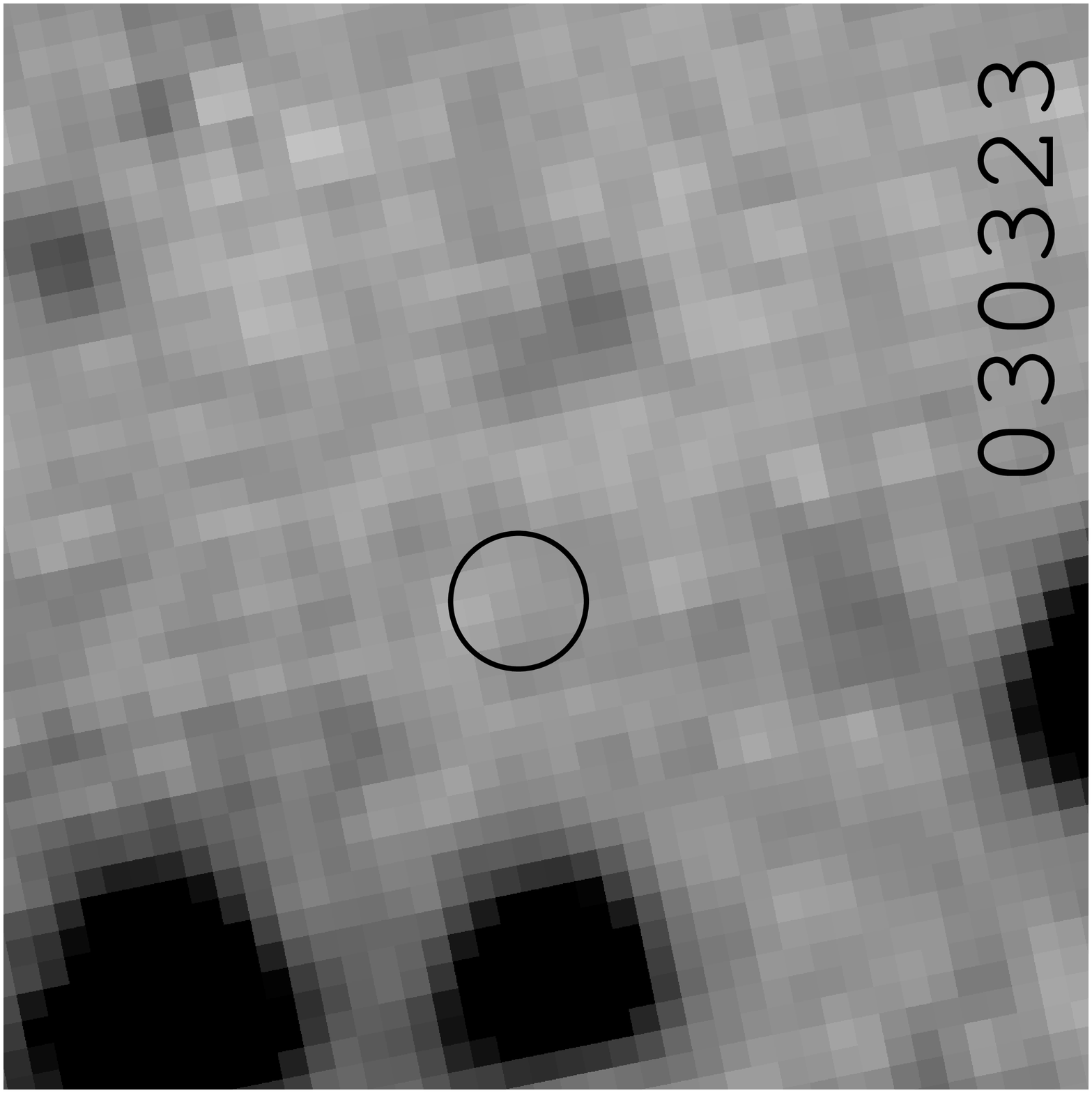}\\
\includegraphics[width=0.24\columnwidth,angle=-90]{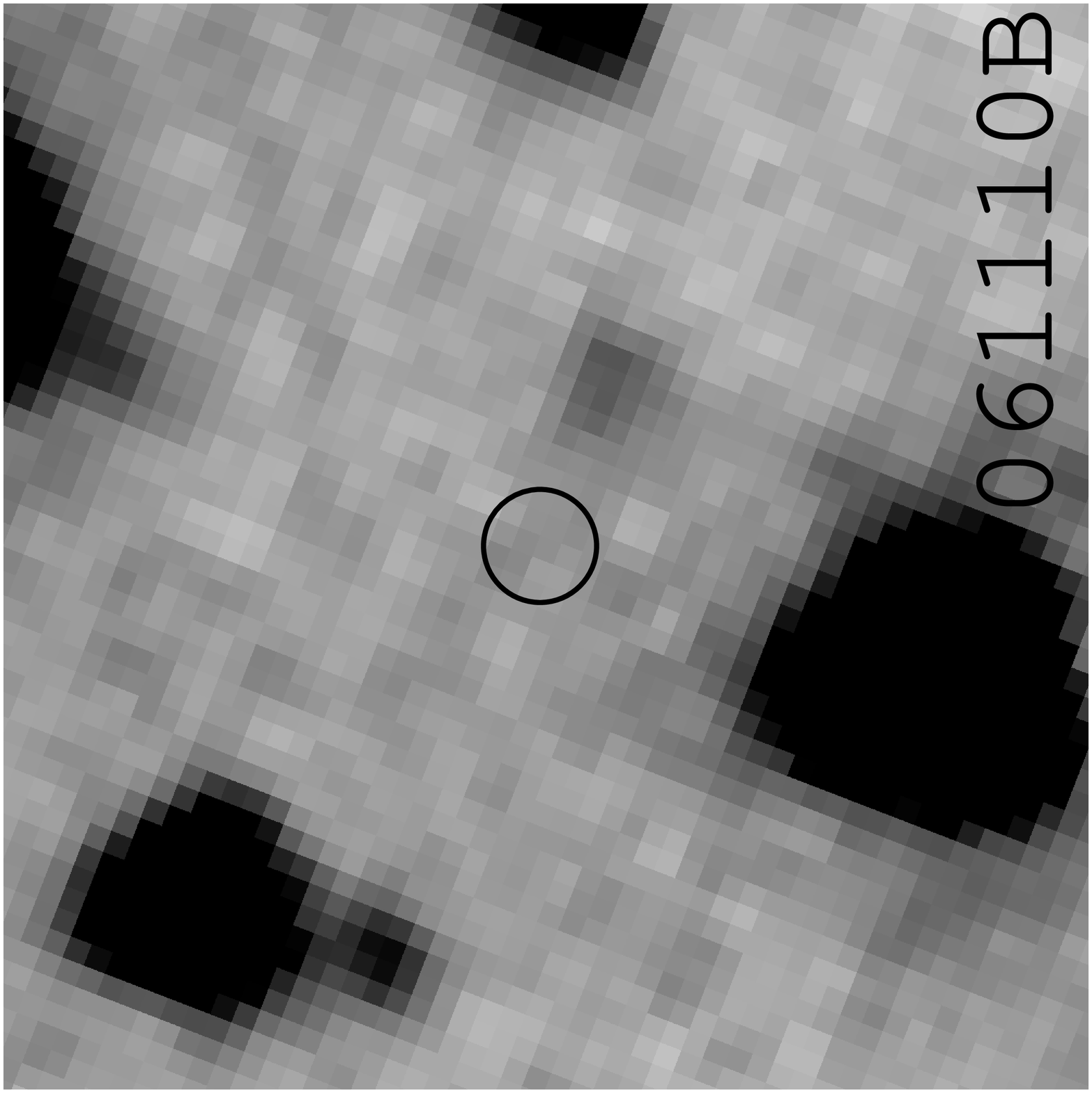}
\includegraphics[width=0.24\columnwidth,angle=-90]{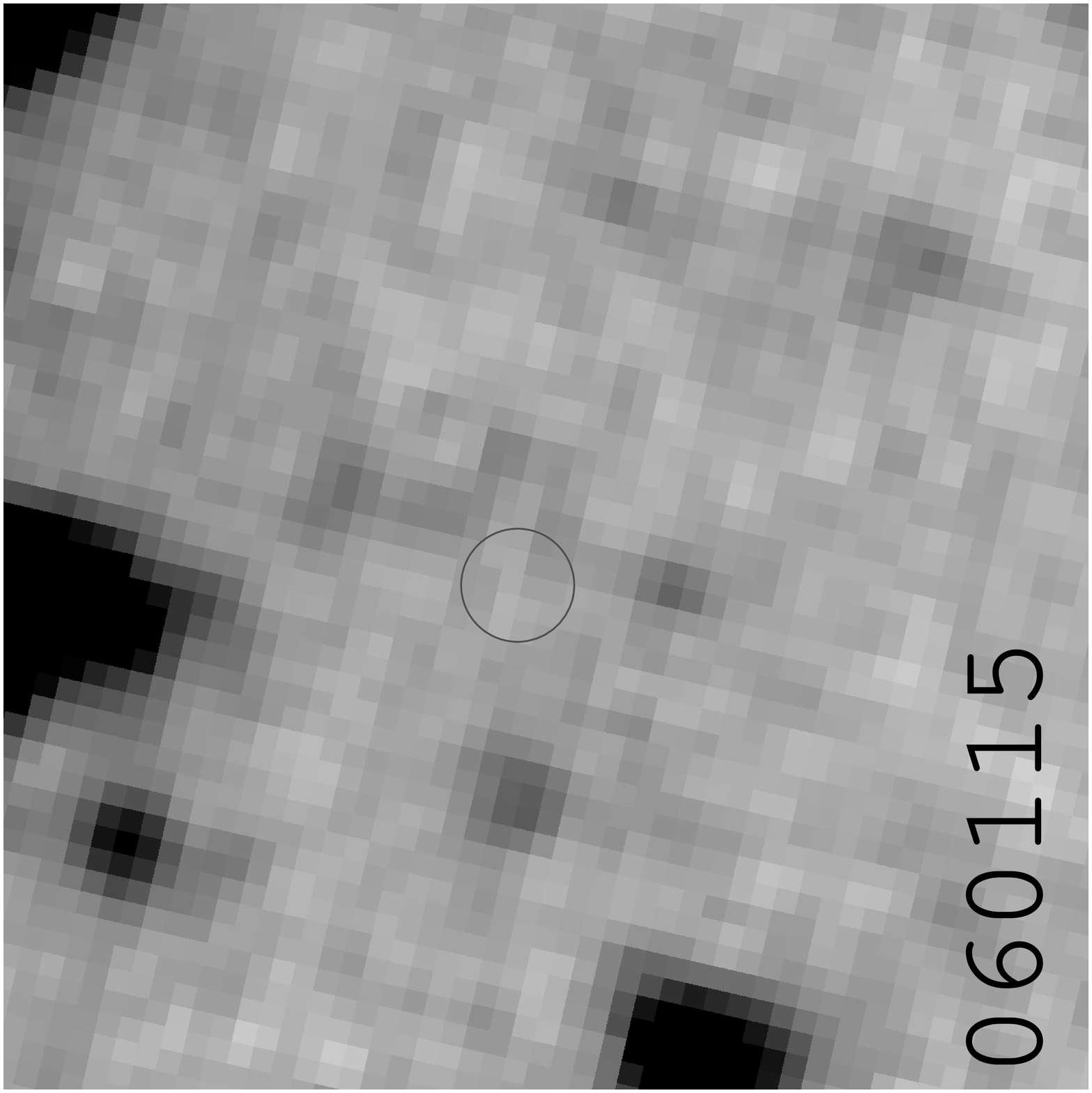}
\includegraphics[width=0.24\columnwidth,angle=-90]{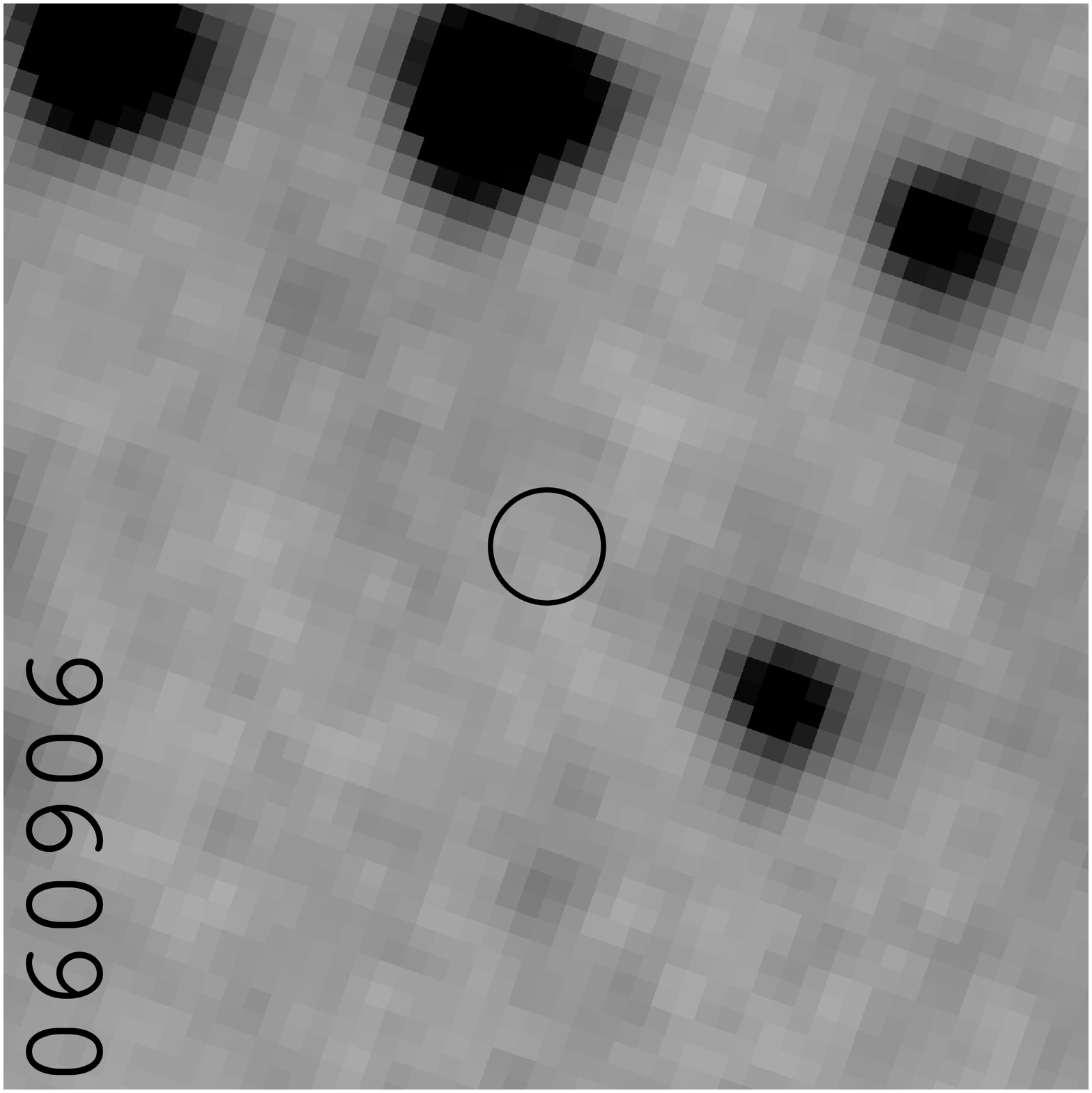}\\
\includegraphics[width=0.24\columnwidth,angle=-90]{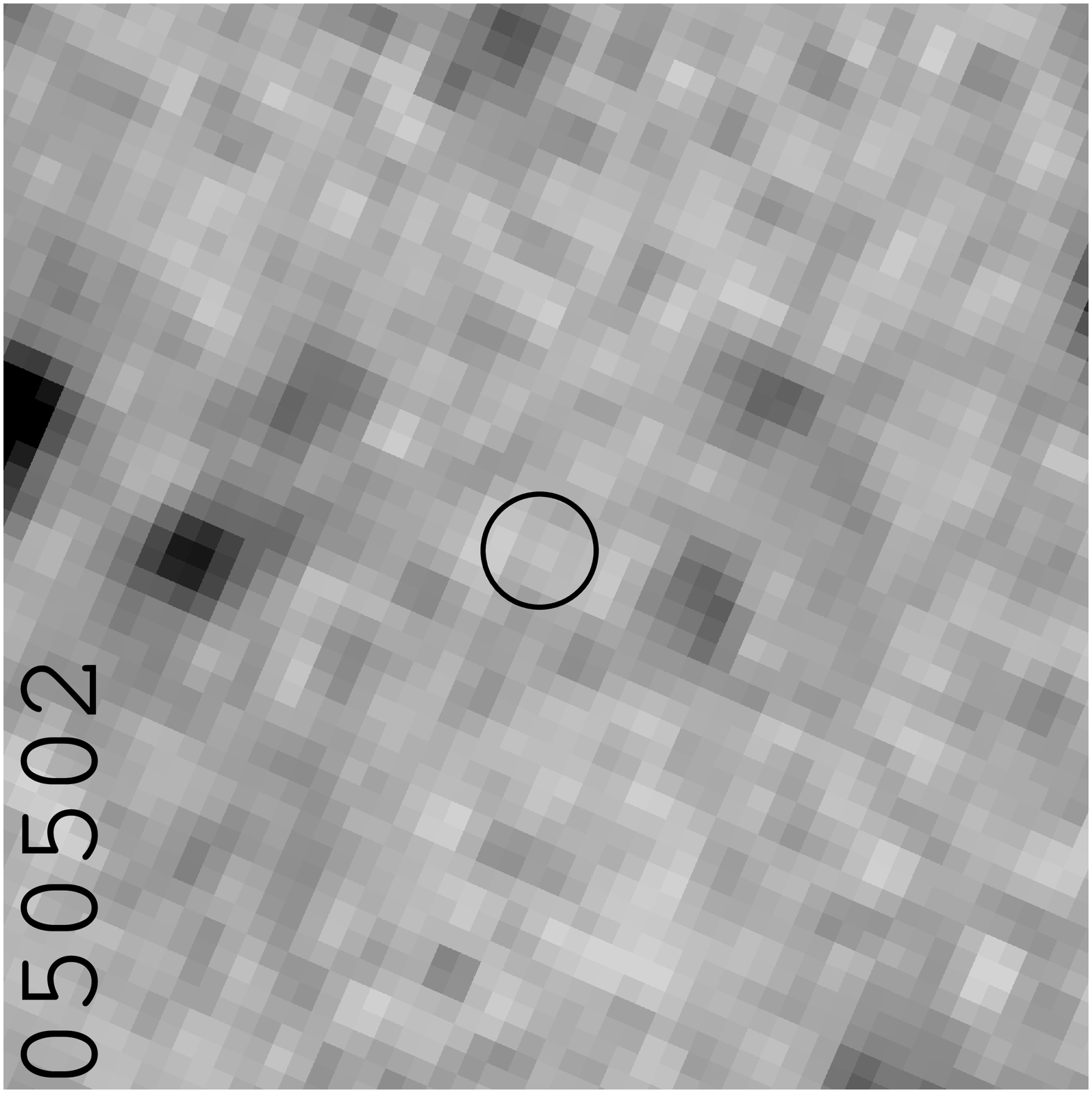}
\includegraphics[width=0.24\columnwidth,angle=-90]{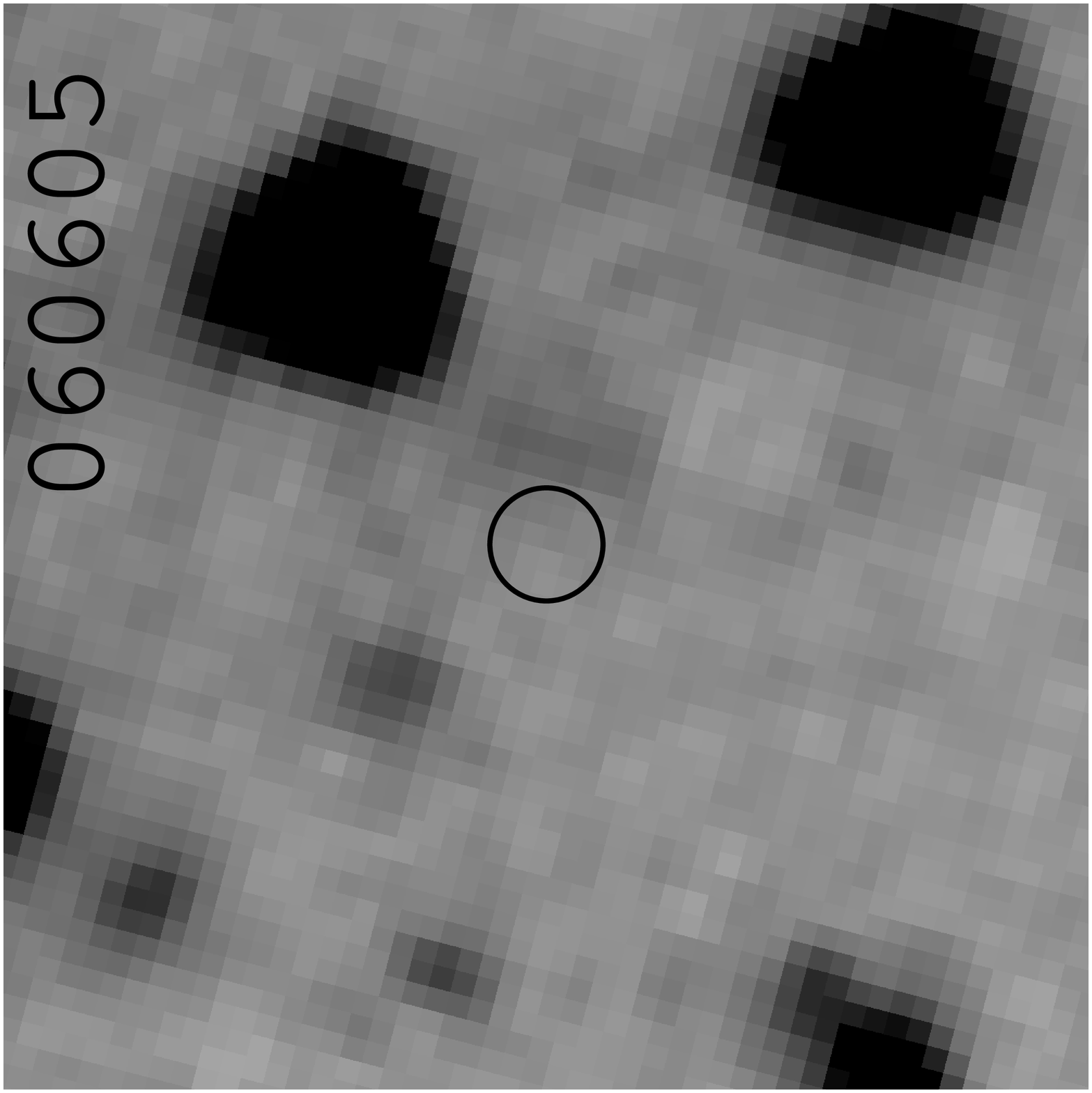}
\includegraphics[width=0.24\columnwidth,angle=-90]{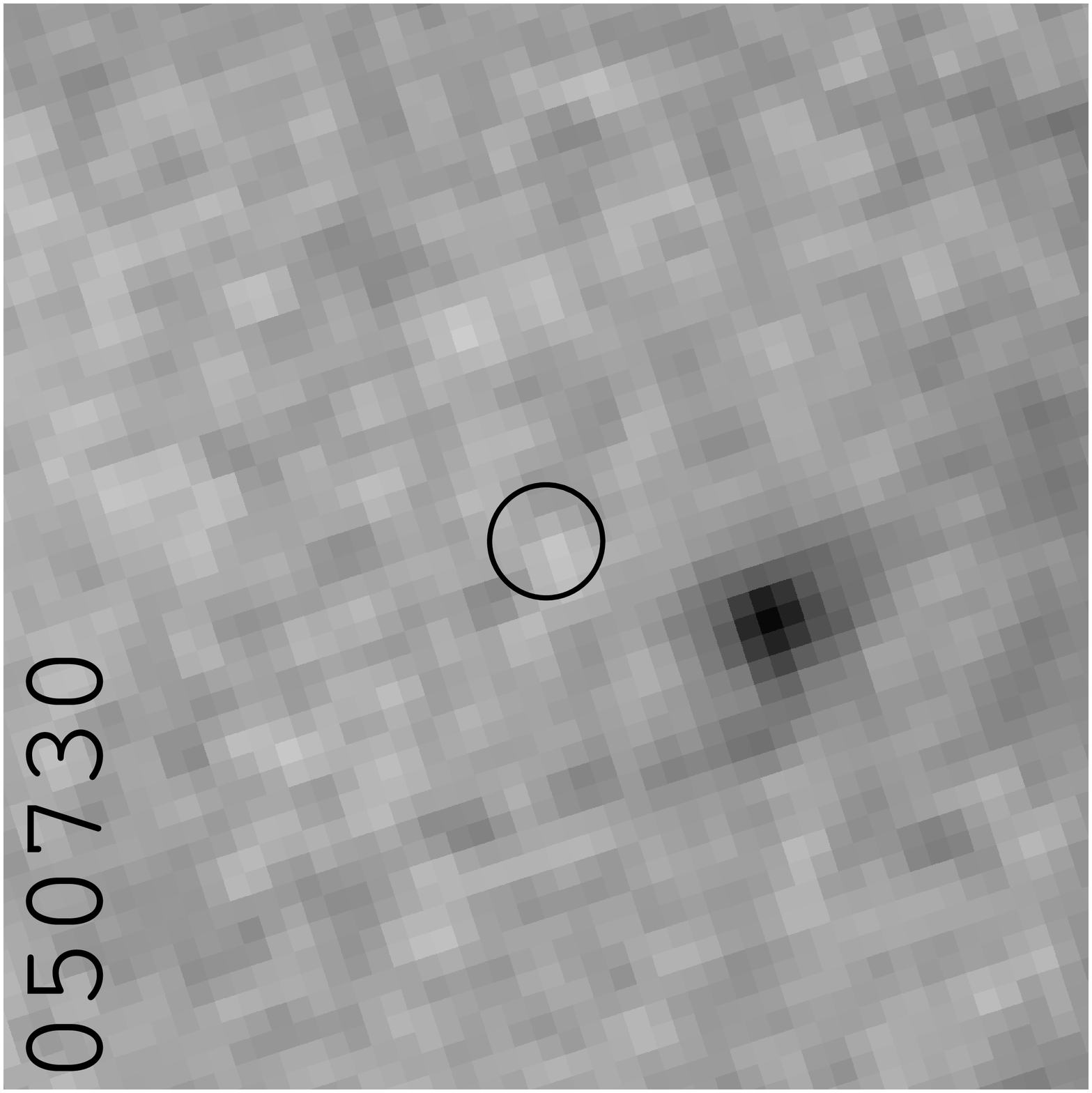}\\
\includegraphics[width=0.24\columnwidth,angle=-90]{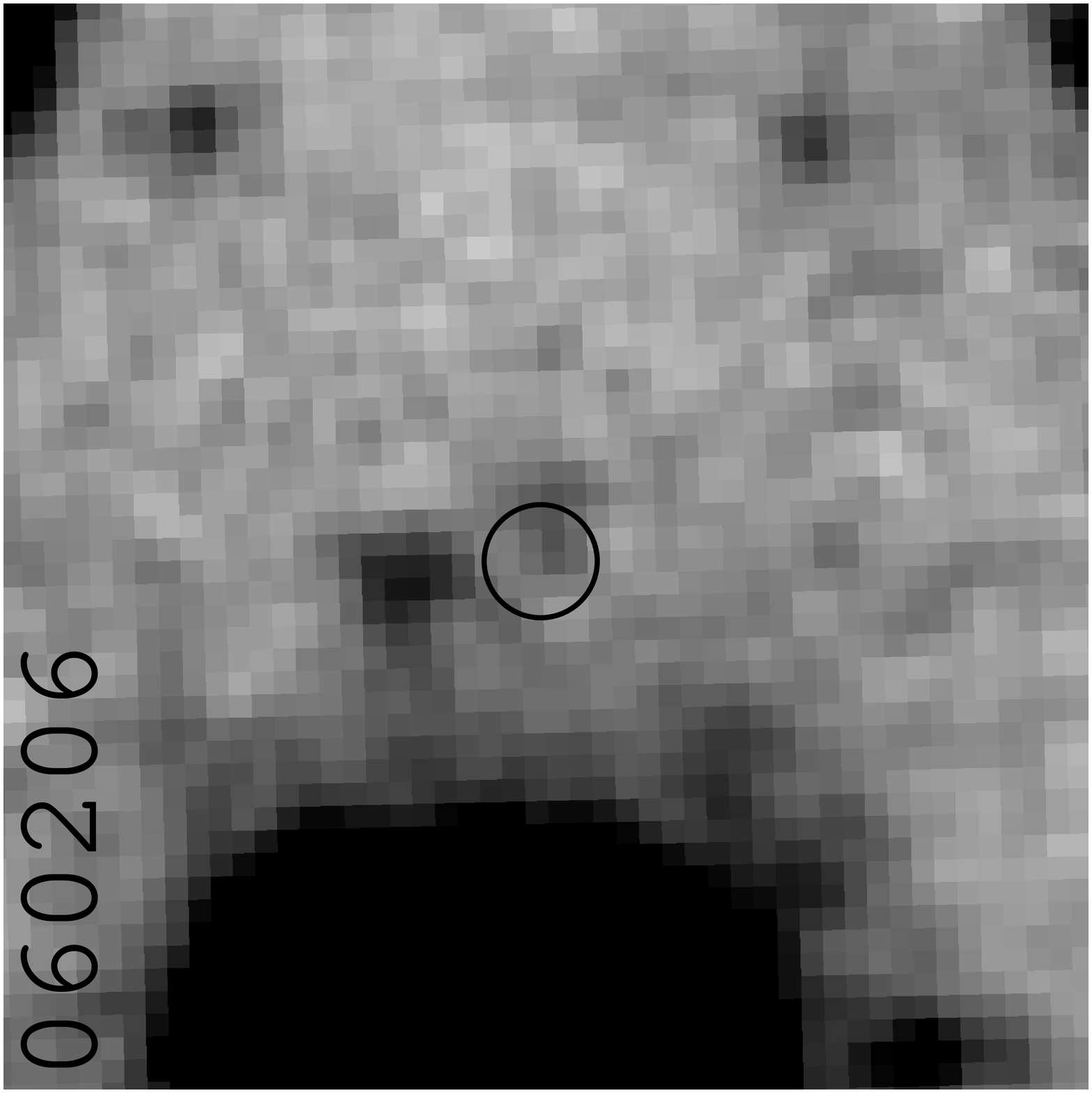}
\includegraphics[width=0.24\columnwidth,angle=-90]{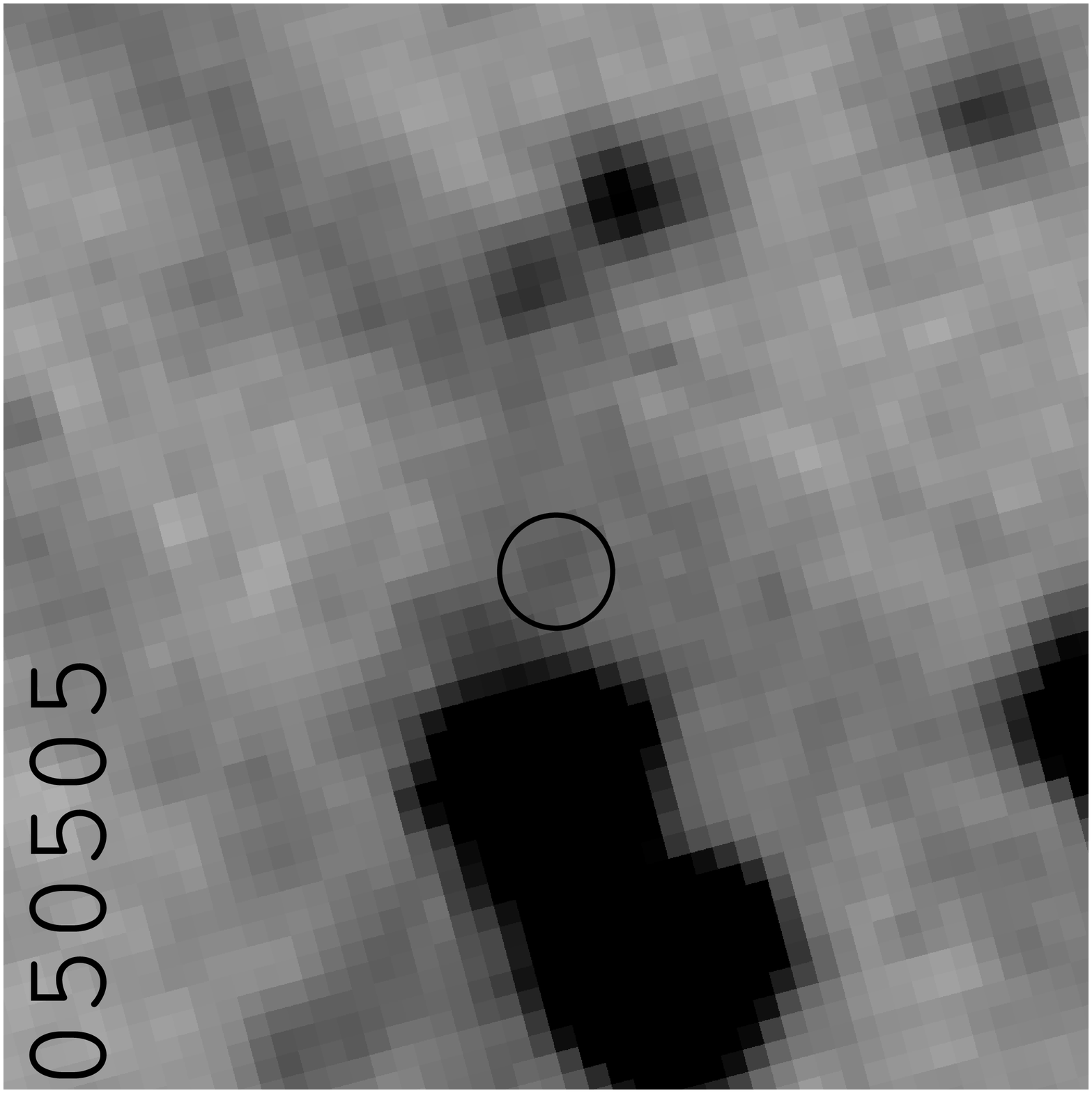}
\includegraphics[width=0.24\columnwidth,angle=-90]{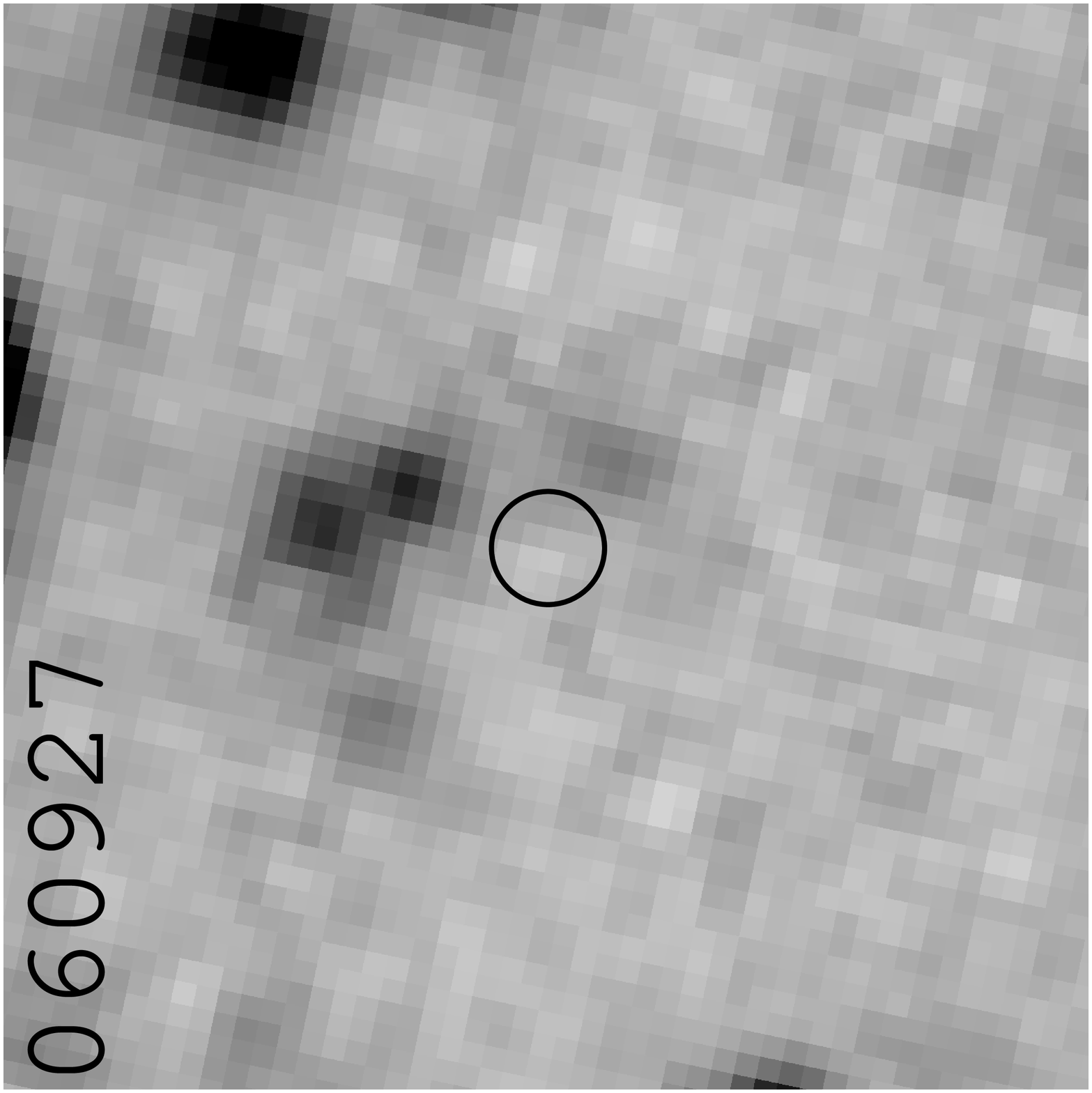}
\end{center}
\caption{\spitzer\ images of regions around the GRB hosts with 3.6
\mum\ non-detections.  The circles (1\arcsec\ radius) mark the
afterglow positions.  All images have the same orientation (North is
up and East is to the left) and scale (16\arcsec\ on a side) with
0\arcsec.4 square pixels.
\label{fig:cutouts2}}
\end{figure}

We used optical afterglow images to perform relative astrometry on the
\spitzer\ mosaics and to locate the GRB hosts.  The median
root-mean-square residual of the astrometric ties is about $0''.12$,
corresponding to about one-tenth of the \spitzer\ PSF at 3.6 \mum.
This is the dominant source of uncertainty in the astrometry, since
the optical afterglow detections themselves are mostly of high
signal-to-noise.  In only the two cases (GRBs 050502 and 050814),
where no afterglow images were available, we performed absolute
astrometry based on SDSS and 2MASS using the afterglow coordinates as
reported in the GCN circulars \citep{Jensen2005,Blake2005}.  We detect
one of these hosts (GRB\,050814) in our \spitzer\ follow-up.  The
\spitzer\ images for the detected hosts are presented in
Figure~\ref{fig:cutouts1}, while non-detections are presented in
Figure~\ref{fig:cutouts2}.

\subsection{Photometry}

At the depth of our observations, \spitzer\ images are
confusion-limited for faint sources.  As a result, in several cases
the region around the expected location of the GRB host is
contaminated by light from nearby stars or galaxies.  Prior to
performing photometry, we used the \textsc{galfit} software package
\citep{phi+02} to model and subtract these neighboring sources.  For
this purpose, we used multiple point sources to determine the mosaic
PSF with the IDL \textsc{starfinder} routines \citep{Diolaiti2000}.
The accuracy of the generated PSF was evaluated by fitting and
subtracting point sources at various locations on the mosaic.  To
remove neighboring sources around the expected location of the hosts,
we used \textsc{galfit} with point-source, Gaussian, or Sersic models as
appropriate in order to achieve the lowest level of residuals. In two
cases (GRBs 061222B and 050505), the expected location of the host
based on the afterglow astrometry fell on the diffraction spike of a
saturated star.  Since we cannot model the PSF at the required level
of accuracy to robustly subtract these diffraction spikes, we do not
consider these two sources in the subsequent analysis. 

For the remaining 18 targets, we searched within $1\arcsec.5$ of the
afterglow centroid (corresponding to $\sim$ 10 kpc at z$\sim3.5$)
and detected 5 GRB hosts at 3.6 \mum\ (GRBs 050319, 050814, 060707, 
060210, and 060926). None of these hosts were detected in the 
simultaneously-observed 5.8 \mum\ IRAC band, which has substantially
worse sensitivity. In the two cases (GRBs 060926 and 060210) where
a nearby source was subtracted prior to photometry, we found
(based on the level of residuals) that the flux we associate 
with the GRB host cannot be explained by modelling it as part 
of the subtracted source. For GRB 050908, a visual inspection 
reveals a coincident flux excess, but photometry indicates 
that it is consistent with a noise fluctuation. For GRB 0600206,
the source $\sim 1\arcsec.5$ to the West of the afterglow centroid
is an unrelated foreground object.

To estimate the probability that one or more of 
the detected sources are chance superpositions, we ran 
\textsc{starfinder}'s source-detection routines on $3'.4\times3'.4$ 
pixel cutouts of the field around our targets and searched for 
sources down to $5\sigma$ (the significance level of the faintest
detection) using the PSF generated from the corresponding images. 
Based on the mean number of sources detected at different thresholds
and following \citet{Bloom2002}, we assign a false detection 
probability given by $P_{\rm cc} = 1-e^{-\pi R^2 \Sigma(u)}$ 
to each of our detections (Table~\ref{tab:data}).
Here $P_{\rm cc}$ is the probability of chance coincidence, 
R is the aperture radius and $\Sigma(u)$ is the number of sources
per unit area down down to the threshold, $u=k\sigma$.
The probability that all of our detections are chance coincidences
is negligible ($10^{-6}$), while the probability that
none of the targets are chance superpositions is 72\%.
Thus, whereas it is possible for one or two of our detections 
to be chance superpositions, it is highly unlikely to be the case for all.

We use the \textsc{funtools} package to perform aperture photometry on
our detections by placing apertures of 2 native IRAC pixels ($2.45''$)
in radius and background annuli of $2-6$ native pixels ($2.45-7.34''$)
in radius centered on the detected sources.  We choose these values
since they allow us to apply standard IRAC aperture
corrections\footnotemark\footnotetext{See section 4.10 of the IRAC
Instrument Handbook.}, which are relevant for the expected compact
sizes of galaxies at $z\gtrsim 3$.  In the cases where this choice of
radii caused nearby objects above the $3\sigma$ level to fall within
either the aperture or the background annulus, we mask them out and
correct for the lost flux (in both the aperture and annulus) by
determining our own aperture correction using mosaics of the IRAC
calibration star HD1812095, prepared with identical parameters as for
our targets.

We determine uncertainties on our measured flux densities using the
uncertainty mosaics created by \textsc{mopex}.  We carry out aperture
photometry in an identical fashion on the (squared) uncertainty images
as for the source images themselves, including aperture corrections as
described above. In addition, we account for correlated noise due to
the drizzling process by incorporating an estimate for it in the flux
density uncertainty\footnotemark\footnotetext{See
http://wise2.ipac.caltech.edu/staff/fmasci/ApPhotUncert\_corr.pdf for
a derivation.}:
\begin{equation}
\sigma_{\rm{src}}^{2} = \ A F_{\rm{corr}}\left[\sum_{i=1}^{N_{\rm{A}}}
\sigma_{i,\rm{A}}^{2} + \frac{N_{\rm{A}}^{2}}{N_{\rm{B}}^{2}}
\sum_{i=1}^{N_{\rm{B}}}\sigma_{i,\rm{B}}^2\right],
\end{equation}
where $\sigma_{\rm{src}}$ is the variance of the source flux density,
$A$ is the aperture correction,
$F_{\rm{corr}}$ is the effective number of pixels over which noise
is correlated in the mosaic, ${N_{\rm{A}}}$ and ${N_{\rm{B}}}$ are the 
number of pixels in the aperture and background region, respectively, and
$\sigma_{i,\rm{A}}$ and $\sigma_{i,\rm{B}}$ are the uncertainty of the
flux density in the $i^{\rm{th}}$ pixel of the aperture and background
regions, respectively.  Since our final mosaic has 0\arcsec.4 pixels,
whereas the native detector pixels are 1\arcsec.22 on a side, 
noise will be correlated over about 3 pixels in our images
(the exact correlation function will depend on the drizzling parameter
as well, for which we used a value of 0.7).
As a conservative estimate, we take $F_{\rm{corr}}$ to be 3.

The resulting flux densities and upper limits are listed in
Table~\ref{tab:data}.  The detections range from about 0.55 to 1.65
$\mu$Jy, while the typical upper limit is about 0.25 $\mu$Jy
(3$\sigma$).  We list the \spitzer\ 3.6 \mum\ results for five
additional GRB hosts at $z\gtrsim 3$ from the literature
\citep{Berger2007a,Chary2007a,Chen2010} in Table~\ref{tab:other}.

\subsection{Stacking}

\begin{figure}[ht]
\centering
\includegraphics[width=0.6\columnwidth,angle=-90]{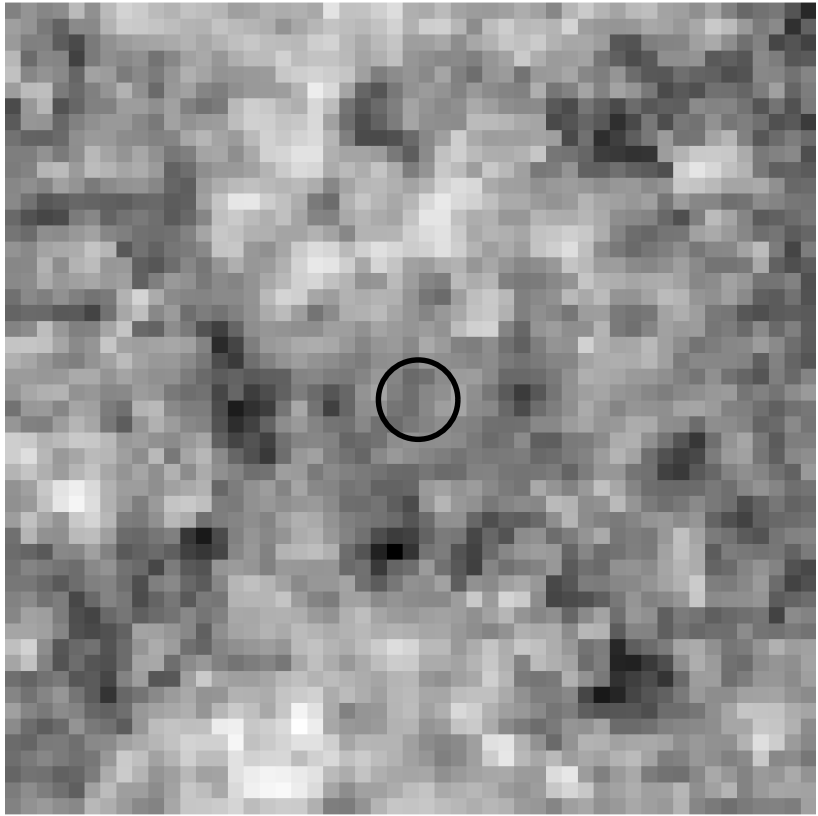} 
\caption{Weighted mean stack of a $51\times 51$ pixel region around
the location of the 11 GRB hosts with precise astrometry and
individual non-detections.  The circle marks a 1\arcsec\ radius
centered on the expected stack location of the hosts.  The
non-detection in the stack yields a 3$\sigma$ upper limit of 80 nJy.
\label{fig:stack}}
\end{figure}

To assess the typical flux density of the non-detected hosts we carry
out a stacking analysis with 11 of the 13 non-detections that have
accurate relative astrometry.  We exclude GRB\,050502, for which we
only have absolute astrometry, and GRB\,060927, for which the relative
astrometry is poor due to a low signal-to-noise detection of the
afterglow.  The remaining 11 targets are located at a median redshift
of $z=3.4$.  We first perform sub-pixel shifts on the \spitzer\
mosaics to bring the expected location of each host (based on the
afterglow centroid) to the center of a mosaic pixel using the IRAF
task \textsc{imshift}.  We then average $51\times 51$ pixel sections
from each image centered on the expected location of the host,
weighted by the inverse of the corresponding variance maps,
after masking out the bright ($>5\sigma$ sources).  The
resulting stacked image (Figure~\ref{fig:stack}) does not show a
detection, and we place a limit on the mean flux density of these 11
hosts of $\lesssim 80$ nJy (3$\sigma$).

\subsection{Metallicities}

Absorption spectra of GRB afterglows present a unique opportunity to
measure ISM abundances of galaxies at $z\gtrsim 3$, where current 
spectroscopic sensitivity limits are inadequate for measuring
metal abundances. A typical optical afterglow spectrum 
exhibits a wide range of ISM absorption features
due to rest-frame UV transitions of low- and high-ionization metal
species, which allow a direct determination of the column density of
these elements along the GRB line of sight through the host galaxy.
Combined with a determination of the neutral hydrogen column density
via the Ly$\alpha$ line, it is possible to determine the ISM
abundances (e.g., \citealt{Berger2006,Prochaska2007,Fynbo2009}).
 
Of the ions typically present in an afterglow spectrum, many are due
to refractory elements and therefore depletion onto dust precludes
their use as robust abundance indicators (they can be used to place a
lower limit on the metallicity).  In this work we use \ion{S}{2}, when
available, as a measure of the metallicity, primarily since sulfur is
not strongly depleted onto dust.  Furthermore, the \ion{S}{2}$\lambda
\lambda 1250.6, 1253.8$ transitions have low oscillator strengths, and
the corresponding lines are more likely to be unsaturated.

In Table~\ref{tab:Z} we present a compilation of metallicities for our
GRB host sample, including the spectral line used, the neutral hydrogen 
column density, and the column density of the metal ion computed using the
measured equivalent width of the transition.  All except one of these
values are taken from the literature
\citep{Hjorth2003,Vreeswijk2004,Savaglio2006,Ferrero2009,Fynbo2009,Ledoux2009,Thone2010}
and have been placed on the solar abundance scale of
\citet{Asplund2005}.  Where an \ion{S}{2} line was not detected, we
use the errors reported on the equivalent widths of transitions in the
vicinity of rest-frame 1250\,\AA\ to place $3\sigma$ upper limits on
the metallicity.  We also place lower limits on the metallicity using
\ion{Si}{2}, \ion{Si}{4} and \ion{C}{2} detections reported by \citet{Fynbo2009}.
Finally, for five GRBs in
our sample, metal line equivalent widths are not available and we do
not consider them in our mass-metallicity analysis (\S\ref{sec:mz}).  The same
quantities for the five additional GRB hosts at $z\gtrsim 3$ from the
literature are listed in Table~\ref{tab:otherZ}.

\section{Optical Luminosities and Stellar Masses of Long GRB Hosts at
$z\sim 3-5$} 
\label{sec:res}

Before we address the $M_*$-$Z$ relation itself, we explore the
rest-frame optical properties of our $z\sim 3-5$ long GRB sample since
this analysis has not been performed previously.  We compare our
sample to long GRB hosts at $z\lesssim 2$ to explore any evolution in
host properties, as well as to field galaxy samples at similar
redshifts to place the long-duration GRB hosts in a broader context.

\subsection{Luminosity Distribution}
\label{sec:lum}

\begin{figure}[ht]
\includegraphics[width=0.48\columnwidth,angle=0]{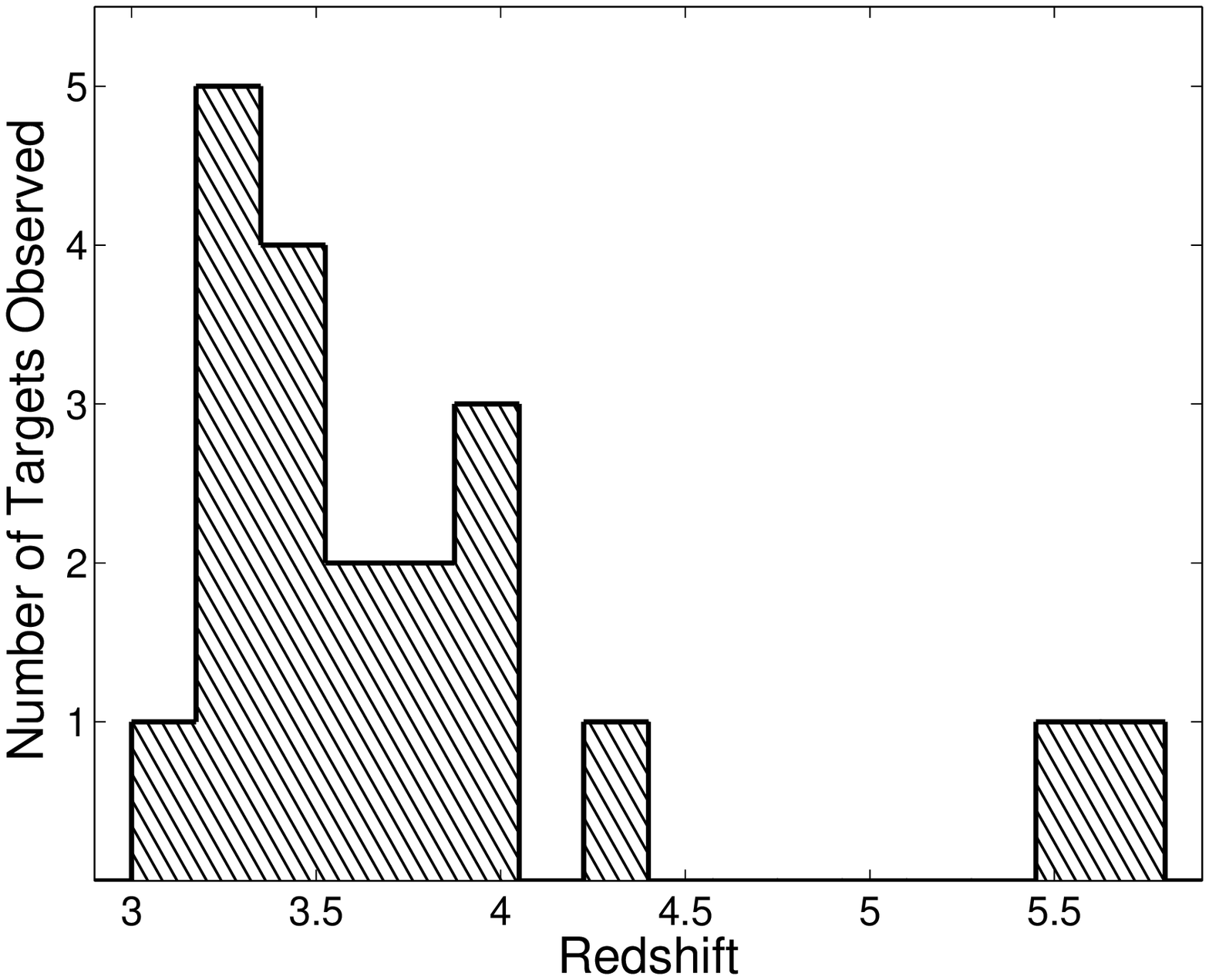}\hfill
\includegraphics[width=0.48\columnwidth,angle=0]{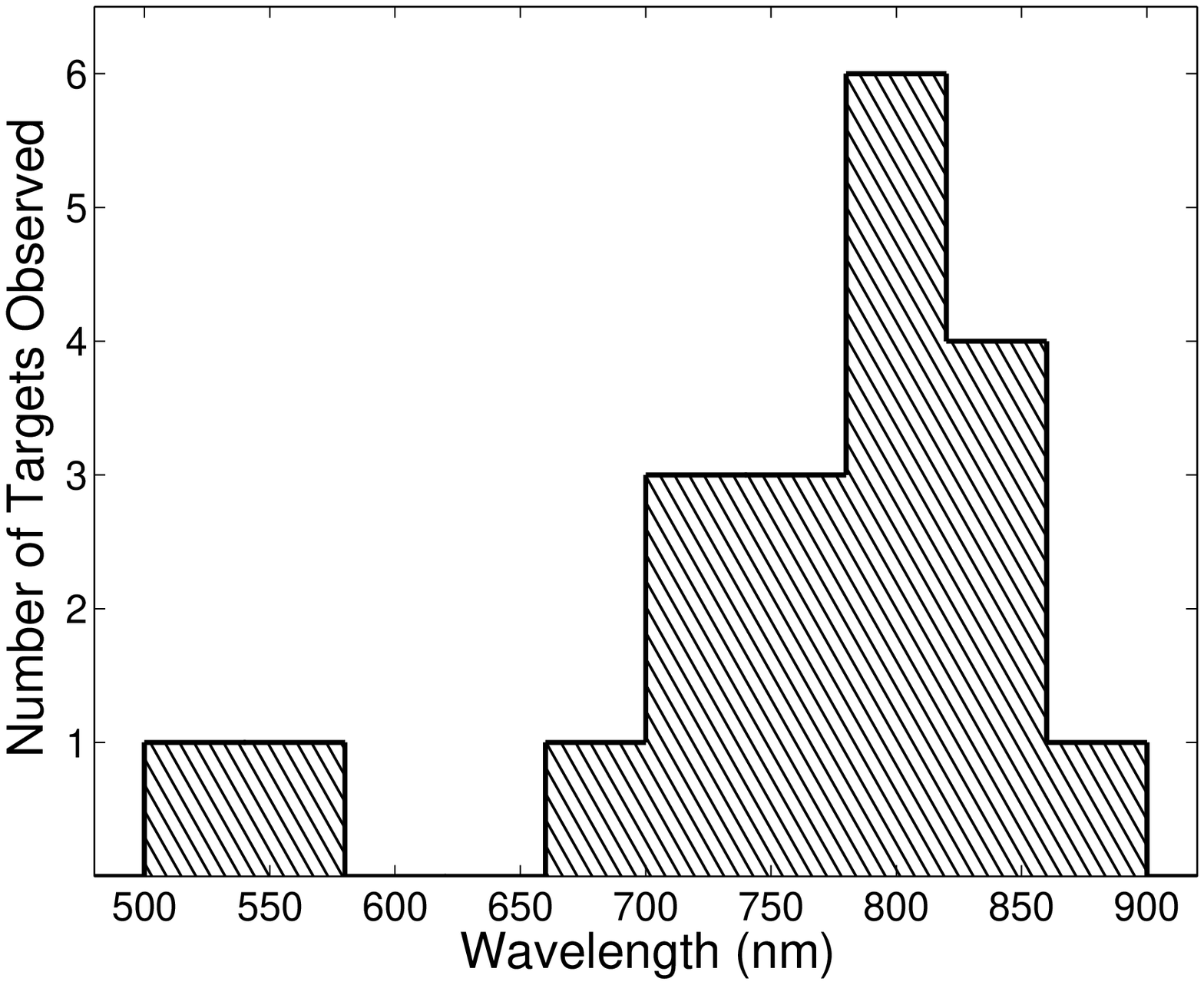}
\caption{Redshift distribution (left panel) and rest-frame wavelength
probed with our \spitzer/IRAC 3.6 $\mu$m observations (right panel)
for the targets in this paper.  All observations fall redward of
4000 \AA\ and therefore probe the rest-frame optical.
\label{fig:z}}
\end{figure}

Since the \spitzer\ 3.6 \mum\ band probes different parts of the host
SEDs at different redshifts (Figure~\ref{fig:z}), we must correct the
inferred luminosities to a common rest-frame wavelength for a
meaningful comparison (K-correction).  Doing so requires knowledge of
the host SED, which we do not have for our targets.  We therefore
employ evolutionary single stellar population (SSP) models 
with a single burst of star formation (e-folding time, $\tau=0$) to
determine the K-corrections.  \citet{Leibler2010} recently performed
stellar population modeling of 23 long GRB hosts at $z\approx
0.03-1.6$ using multi-band photometry from \citet{Savaglio2009} and the
evolutionary models of \citet{Maraston2005}.  They determined a median
stellar population age of $10^{-1.2\pm 0.1}$ Gyr (see also
\citealt{Savaglio2009}).  Taking this age range into account, along with a
Salpeter IMF, and a metallicity range of $0.05-0.5$ Z$_\odot$, we find
that the flux density of the SSP models in the rest-frame $0.4-2$
\mum\ range is well-approximated by a power law, $F_{\lambda} \propto
\lambda^\beta$, with $\beta=-2.3\pm 0.2$.  This assumes negligible
extinction, which is appropriate for most GRB host galaxies
\citep{Savaglio2009}.  Using this result as an estimate of the intrinsic
spectrum of GRB hosts at $z\gtrsim 3$, we determine the host absolute
magnitudes including K-correction as follows:
\begin{eqnarray}
M_{\rm AB}(\lambda_0) &=& m_{\rm{AB}} - 5\log{\left(\frac{d_{\rm
L}}{\mbox{10 pc}}\right)} \nonumber \\
&& - 2.5(\beta+2)\log{\left[\frac{(1+z)\lambda_0}{\lambda}\right]}
\nonumber \\
&& + 2.5\log{(1+z)},
\label{eqn:kcorr}
\end{eqnarray}
where $M_{\rm{AB}}$ is the absolute magnitude at a rest wavelength,
$\lambda_0$, to which the K-correction is performed, $d_L$ is the
luminosity distance, and $\lambda$ is the effective wavelength of the
observed band (3.55 \mum\ for our \spitzer\ observations).  To
minimize the K-corrections for our sample we select a nominal
rest-frame wavelength, $\lambda_0 = 7800$\,\AA\ (Figure~\ref{fig:z}), roughly
corresponding to the $I$-band.  The resulting mean K-correction on our
photometry is about $-1.7$ mag (dominated by the last term on the
right-hand-side of Equation~\ref{eqn:kcorr}), with a standard
deviation of about 0.2 mag.

\begin{figure}[ht]
\centering
\includegraphics[width=1.0\columnwidth]{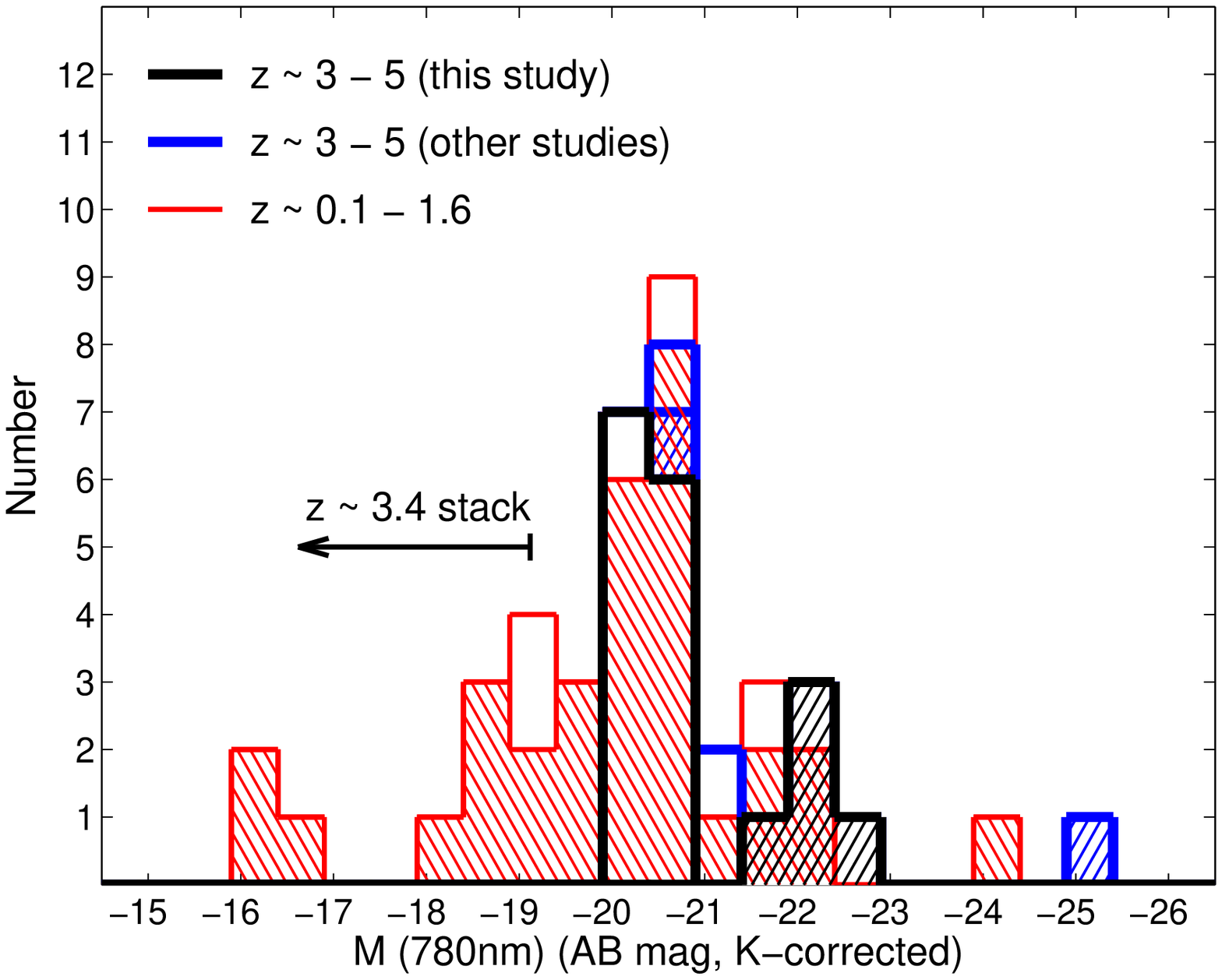}
\caption{The luminosity distribution for GRB hosts at $z\sim 3-5$ from
our study (black) and the 5 GRB hosts at $z\gtrsim 3$ from previous studies
(blue; \citealt{Berger2007a,Chary2007a,Chen2010}), 
compared with GRB hosts at $z\sim 1$ (red). 
All samples have been K-corrected to 780 nm in the rest frame.
Non-detections (3$\sigma$ upper limits) are shown as open histograms.
Also shown is the stack limit at a median redshift of $z\approx 3.4$.
The host of GRB\,080607 at $M_{780~{\rm nm}}=-25.1$ was 
targeted due to the large extinction inferred from the optical
afterglow \citep{Chen2010}.
\label{fig:Llowz}}
\end{figure}

The resulting luminosity distribution for our sample is shown in
Figure~\ref{fig:Llowz}, and the inferred absolute magnitudes are
listed in Table~\ref{tab:data}. Also included are the 5 host galaxies 
from the literature (their values are listed in Table~\ref{tab:other}).
In the comparison to low redshift GRB hosts and to field galaxies provided
below we treat separately our uniform sample, and the combined sample
that includes the 5 hosts from the literature.  For our detected hosts
we find a range of $M_{AB}(780\,{\rm nm})\approx -21.5$ to $-22.5$
mag, while the limits are typically $M_{AB}(780\,{\rm nm})\gtrsim -20.4$
mag.  The stack limit corresponds to $M_{AB}(780\,{\rm nm})\gtrsim
-19.1$ mag.  With less than a $50\%$ detection fraction it is not
possible to robustly estimate the median luminosity of our sample, but
the formal $3\sigma$ upper limit is about $-21.5$ mag.  The stack
limit, however, suggests that a more reasonable upper bound on the
median is about $M_{AB}(780\,{\rm nm})\gtrsim -19.1$ mag.  The addition
of the 5 hosts from the literature does not change this result, with
the exception of the highly luminous host of GRB\,080607
\citep{Chen2010}, which was specifically targeted due to evidence of
large afterglow extinction.

\subsubsection{Comparison with Long GRB Hosts at $z\lesssim 2$}

To compare the resulting optical luminosity distribution to GRB hosts
at low redshift, we obtain a comparison sample from ground-based $JHK$
and {\it Hubble Space Telescope} 814~nm photometry reported in
\citet{CastroCer'on2008} and \citet{Savaglio2009}, and from our own
GRB host follow-up studies.  The comparison sample has a redshift
range of $z\approx 0.01-2$ with a median of $z\approx 0.86$.  For each
host we select the band that corresponds most closely to a rest frame
wavelength of 7800 \AA\ to minimize the K-corrections relative to our
\spitzer\ sample.  The mean and standard deviation of the
K-corrections are $-0.6$ and 0.3 mag, respectively.

The luminosity distribution for the low redshift sample is shown in
Figure~\ref{fig:Llowz}.  There is clear overlap
between the two GRB host samples at the bright end, but due to the
lower redshifts of the comparison sample, its luminosity distribution
extends to much fainter levels (reaching $\approx -16.5$ mag), with a
median of about $-20.1^{+0.9}_{-0.4}$ mag ($95\%$ confidence range).
This is consistent with the upper bound on the median luminosity of
our $z\sim 3-5$ sample.  Indeed, a log-rank test including the
individual non-detections indicates that the hypothesis that two
samples are drawn from the same underlying populations has a p-value
of $0.65$.  Similarly, the fraction of detected hosts above our
threshold of $M_{AB}(780\,{\rm nm})\approx -21.5$ mag is about 1/4 for
both samples.  Finally, our stack non-detection level of
$M_{AB}(780\,{\rm nm})\gtrsim -19.1$ mag is consistent with the median
of the $z\sim 1$ GRB sample.  Thus, we find no evidence for
significant evolution in the optical luminosity function of GRB hosts
from $z\sim 1$ to $\sim 4$.  We note that dividing the comparison
sample into $z<1$ and $z>1$ subsets does not change this result.

\subsubsection{Comparison with Lyman-Break Galaxies and Lyman-alpha
Emitters}

\begin{figure}[ht]
\centering
\includegraphics[width=1.0\columnwidth]{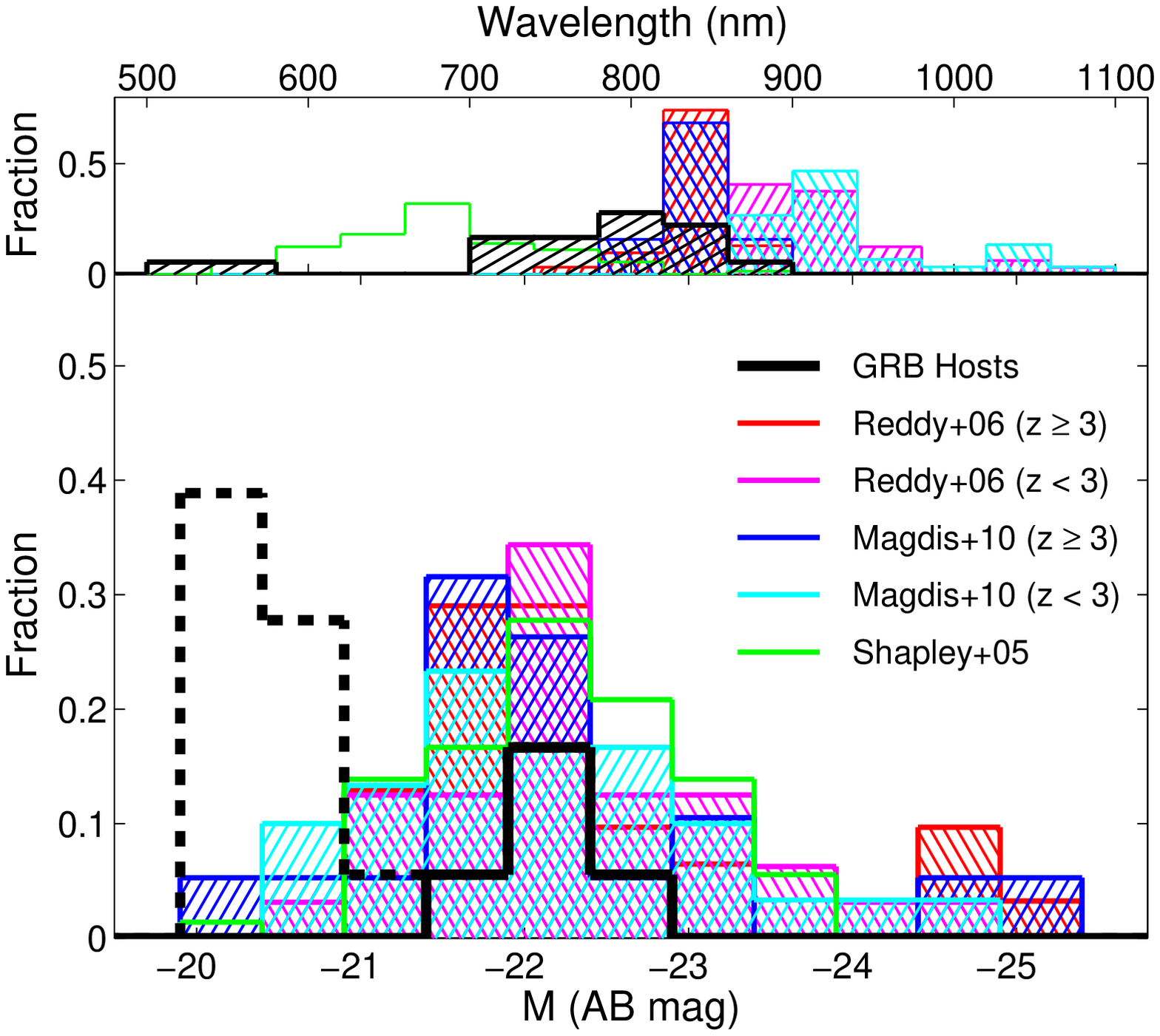}
\caption{Normalized luminosity distribution for the GRB hosts in our
sample (black: solid=detections; dashed=limits) compared with 
LBGs at $z\sim 3$.  \citet{Reddy2006} and
\citet{Magdis2010} use \spitzer\ 3.6 \mum\ observations of GOODS-N 
and probe deeper than our study (Table~\ref{tab:comp}), while 
\citet{Shapley2005} rely on ground-based $K$-band photometry.  
K-corrections for the SED shape (last term in equation \ref{eqn:kcorr}) have not
been applied (although the relative difference
in K-correction should be minor and will not modify the shape of the distribution). 
Non-detections have been removed from the comparison samples for clarity.
The upper panel shows the corresponding distributions 
of the rest frame wavelengths for each sample.  
\label{fig:Lhighz}} 
\end{figure}

To assess whether the luminosities we find for GRB hosts at $z\gtrsim
3$ are typical of field galaxies, we next compare the resulting
luminosity distribution with other galaxy samples at similar
redshifts: Lyman-break galaxies (LBGs) and Lyman-alpha emitters
(LAEs).  For the LBG comparison we use \spitzer\ 3.6 \mum\ photometry
in the GOODS-N field reported by \citet{Reddy2006,Magdis2010}, 
based on deep ($\sim 95$-hour) \spitzer\ observations.
We also include a sample of 72 star-forming
galaxies at $z\sim 2.3\pm 0.3$ from \citet{Shapley2005} selected based
on their rest-frame UV brightness.  The resulting luminosity
distributions for these samples are shown in Figure~\ref{fig:Lhighz},
and their summary statistics are listed in Table~\ref{tab:comp}.
K-corrections for the SED shape (third term on the RHS of equation 
\ref{eqn:kcorr}) have not been applied to any of the samples,
although the difference between the K-corrections should be minor 
($\lesssim0.2$ mag) and would not modify the shape of the distributions.

Our GRB host sample is clearly missing the luminous tail of LBGs at
$M_{\rm{AB}}\lesssim -23$ mag, which accounts for $\approx 20\%$ of
the comparison samples.  Even if we include the luminous host of
GRB\,080607, it accounts for only $4\%$ of the GRB host sample
observed with \spitzer.  The median absolute magnitude of the LBG
sample is $-22.0_{-0.2}^{+0.3}$ mag for \citet{Reddy2006} 
and $-21.9_{-0.2}^{+0.5}$ mag for \citet{Magdis2010} ($95\%$ confidence ranges),
brighter than the 3$\sigma$ upper limit for the GRB sample ($-21.5$ mag, 
even without K-corrections for the SED shape) and our stack
limit.  A log-rank test yields a p-value of $0.23$
and $0.38$ that the GRB host sample is drawn from the same
population as the parent population of the LBG samples of \citet{Reddy2006}
and \citet{Magdis2010}, respectively.
It is important to note, however, that since the LBG sample is flux
limited (based on the initial optical selection and spectroscopic
confirmation), whereas the GRB host sample is not, we cannot simply use
the fractional detections of LBGs as an indication of the overall
luminosity function.

To further assess whether the dearth of GRB hosts with
$M_{\rm{AB}}\lesssim -23$ mag in our sample is significant, we instead
need to integrate the rest-frame optical luminosity function.  This
will allow us to assess the expected fraction of GRB hosts with
$M_{\rm{AB}}\lesssim -23$ mag compared to our threshold of about $-21$
mag.  \citet{mvq+07} calculated a Schechter fit to the $V$-band
luminosity function at $z\sim 3$ and found $M_{\rm AB}^*(V)=-22.77\pm
0.22$ mag and faint-end slope, $\alpha=-1.12\pm 0.24$. 
Applying a K-correction from $V$ to $I$ band using 
$F_\lambda\propto \lambda^{-2.3\pm 0.2}$ (\S\ref{sec:lum}) 
we find $M_{\rm AB}^*(I)=-22.65\pm 0.30$ mag.
Assuming that $\alpha$ is the same in the I-band as in the V-band, 
we find that about $10\%$ of our sample
(or about 2 hosts) should have $M_{\rm{AB}}\lesssim -23$ mag if the
GRB hosts are drawn from the field galaxy
population\footnotemark\footnotetext{We verify this approach by
calculating the expected fraction of galaxies with
$M_{\rm{AB}}\lesssim -23$ mag for our various comparison samples.  We
find that relative to their typical threshold absolute magnitude of
about $-21.5$ to $-22$ mag (Figure~\ref{fig:Lhighz}), this expected
fraction is about 20\%, which is in good agreement with the observed
fraction.}.  Since this small number is fully consistent with zero
detections, we cannot rule out the hypothesis that GRB
hosts are drawn from the general LBG population.

\begin{figure}[ht]
\centering
\includegraphics[width=1.0\columnwidth]{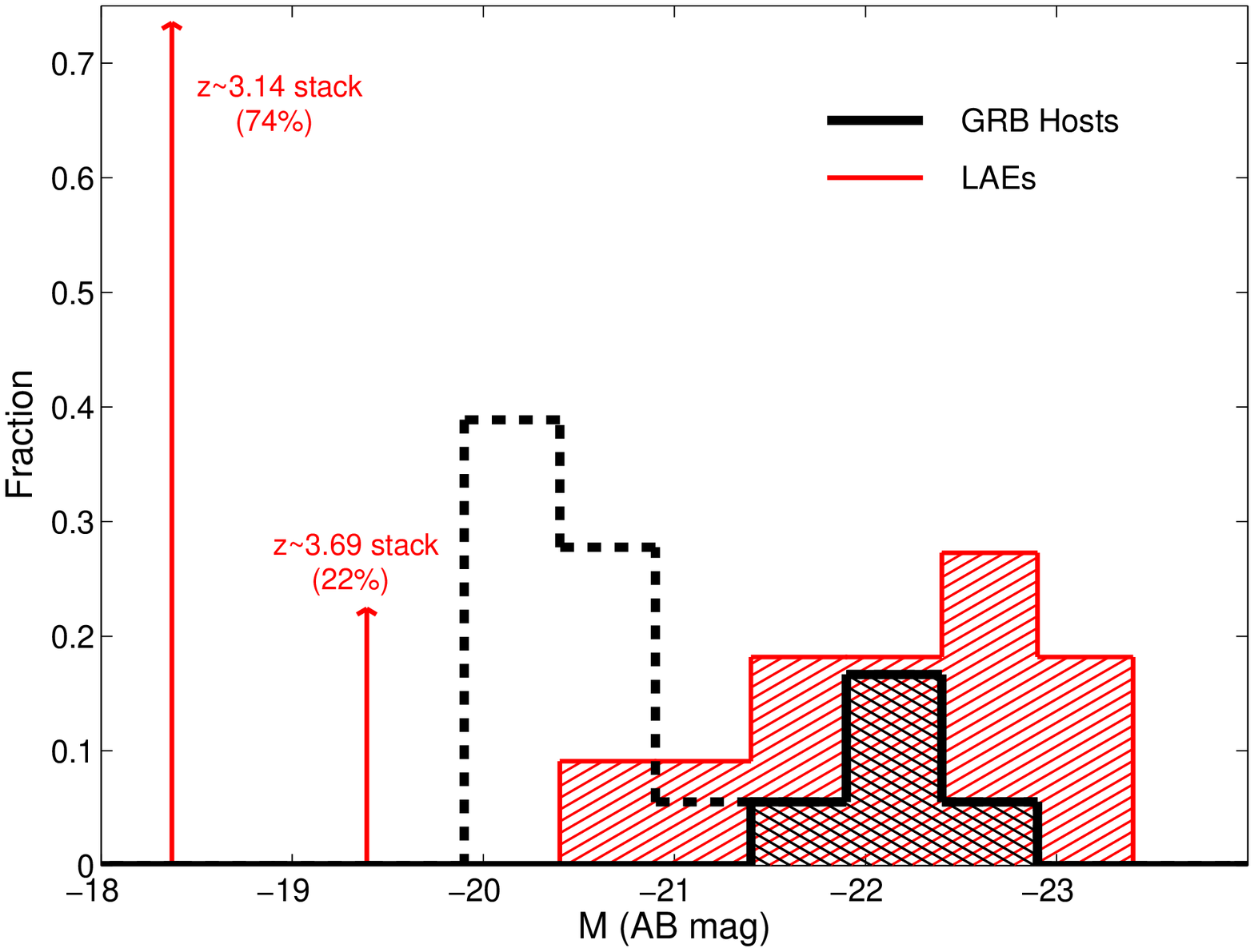}
\caption{Same as Figure~\ref{fig:Lhighz}, but compared with \spitzer\
3.6 \mum\ observations of Lyman-alpha emitters at $z\sim 3.1$ and
$\sim 3.7$ (red; \citealt{Ono2010}).  Only $4\%$ of the LAE sample was
detected individually (red histogram), while stacks at $z\sim 3.1$ and
$\sim 3.7$ revealed much lower typical luminosities of $M_{\rm
AB}\approx -18.3$ mag and $\approx -19.4$ mag, respectively.
K-corrections for the SED shape (last term in equation \ref{eqn:kcorr}) have not
been applied (although the relative difference
in K-correction should be minor and will not modify the shape of the distribution). 
\label{fig:Llae}} 
\end{figure}

For the comparison to Lyman-alpha emitters we use the sample of
\citet{Ono2010}, which includes 205 LAEs at $z\sim 3.1$ and 67 LAEs at
$z\sim 3.7$ with multi-band photometry.  These authors find 11
detections at 3.6 \mum\ (5 at $z\sim 3.1$ and 6 at $z\sim 3.7$),
corresponding to a detected fraction of only $4\%$.  From a stacking
analysis of the non-detected LAEs they determine $\langle M_{\rm AB}
\rangle=-20.8$ mag at $z\approx 3.1$ and $\langle M_{\rm AB}\rangle
=-21.1$ mag at $z\approx 3.7$.  A comparison between the GRB hosts and
LAEs at $z\sim 3.1-3.7$ is shown in Figure~\ref{fig:Llae}.  The
luminosity distributions of the detected LAEs and GRB hosts appear to
be consistent.  However, including the LAE non-detections and carrying
out a log-rank test we find a negligible probability that the GRB host
sample is drawn from the same population as the LAE sample at $z\sim
3.4$ since our sample has a much higher detection fraction than the
LAEs.  This is an interesting result since several GRB hosts have been
previously detected as LAEs (e.g., \citealt{Fynbo2002,Fynbo2003,jbf+05}).
It suggests that GRBs select the more luminous end of the LAE
luminosity distribution.

To conclude, the comparisons to LBGs and LAEs suggest that GRB hosts
at $z\gtrsim 3$ are currently missing the bright end of the LBG
luminosity distribution (with the exception of the host of
GRB\,080607), but that this may be due to the small sample size.  On
the other hand, GRB hosts sample the high end of the LAE luminosity
distribution.

\subsection{Stellar Mass Distribution}
\label{sec:mass}

We next turn to a derivation of the stellar masses of our GRB host
sample.  Computing stellar masses from observed luminosities in a
given wave-band requires knowledge of the mass-to-light ratio and
hence the stellar population age and metallicity.  When multi-band
photometry is available, modeling of the spectral energy distribution
(SED) using stellar population synthesis models can used to determine
stellar masses, provided that a single stellar population is assumed.
When multi-band photometry is not available, the resulting uncertainty
in the mass-to-light ratio (e.g., at $\sim 1$ \mum) is about an order
of magnitude (e.g., \citealt{Magdis2010}).

\begin{figure}[ht]
\centering
\includegraphics[width=1.0\columnwidth]{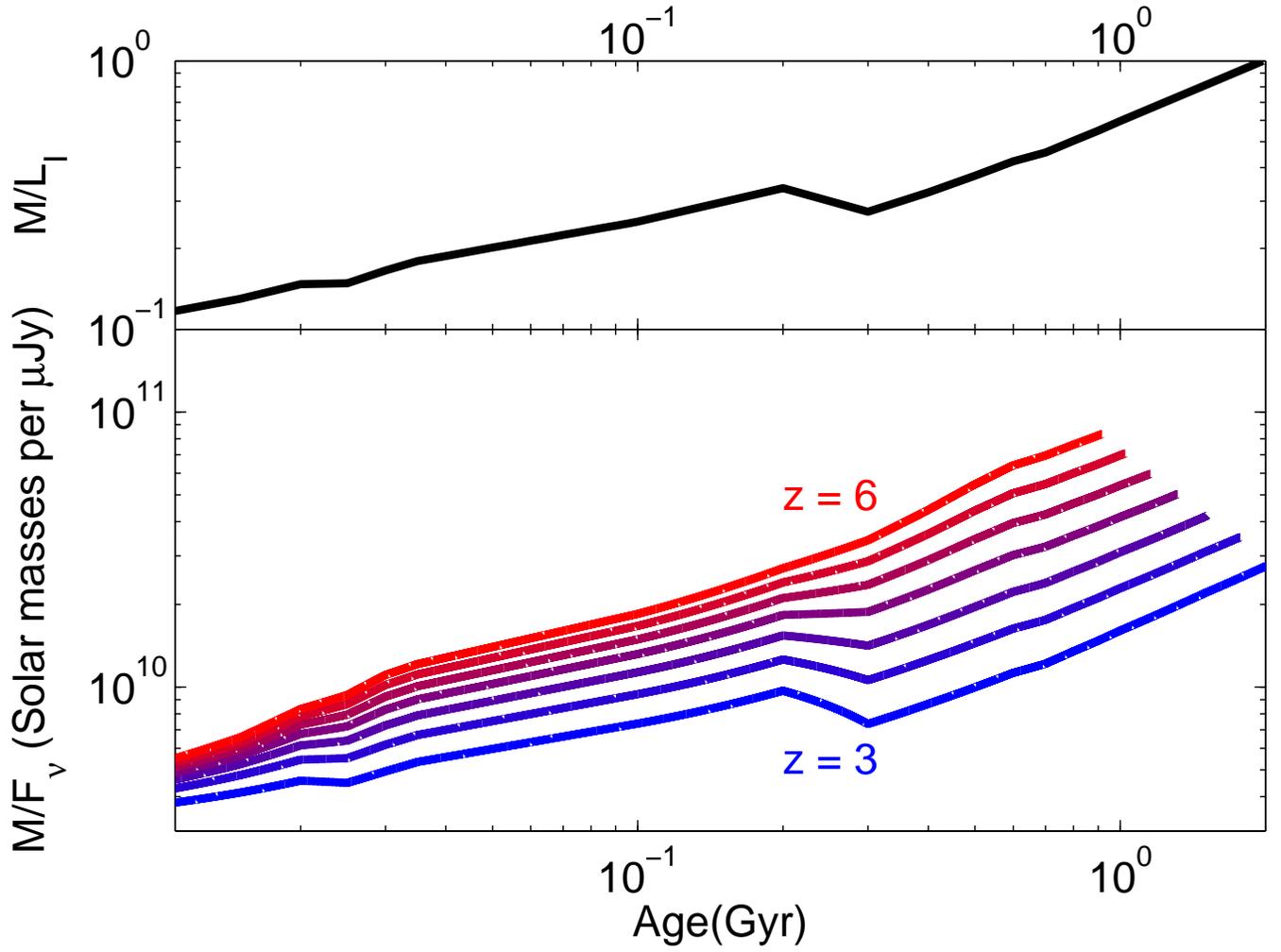}
\caption{Upper panel: mass-to-light ratio in the rest-frame $I$-band
from the \citet{Maraston2005} simple stellar population models in
Solar units. The models assume an instantaneous burst of star formation ($\tau=0$).
Lower panel: ratio of stellar mass to observed
3.6 \mum\
flux density for $(\tau=0)$ SSP models as a function of age at different
redshifts from $z=3$ (bottom curve) to $z=6$ (top curve) in steps of
$\delta z=0.5$.  The models used in this paper are for a Salpeter IMF
with a red horizontal branch morphology and a metallicity of $0.02
Z_{\odot}$.  Each curve is truncated at a value that corresponds to
the age of the universe at that redshift.
\label{fig:ml}}
\end{figure}

Here, since we lack broad-band photometry, we determine a range of
mass-to-light ratios for each galaxy in the observed 3.6 \mum\ band
using a wide range of population ages and the single stellar
population models of \citet{Maraston2005} with a Salpeter IMF. 
We assume an instantaneous burst of star formation ($\tau=0$). As
expected, the 3.6 \mum\ mass-to-light ratio for these models increases
with stellar population age beyond $\sim 10$ Myr.  This is shown in
Figure~\ref{fig:ml}, where the ratio of the stellar mass to the
observed flux density at 3.6 \mum\ is plotted as a function of age and
redshift.  The more traditional mass-to-light ratio in solar units (in
the rest-frame $I$-band) is also plotted for comparison.  The upper
bound on the mass-to-light ratio is achieved by setting the stellar
population age to the age of the universe at each host redshift
($\approx 1.8$ Gyr at the median redshift of our sample).  We stress
that this leads to a very conservative maximum mass for each host
galaxy since studies of LBGs and LAEs indicate typical population ages
of $\sim 0.1-0.6$ Gyr
\citep{Shapley2005,Reddy2006,Magdis2010,Ono2010}.  For a more typical
age we adopt the median age for long GRB hosts at $z\sim 1$ of about 70 
Myr \citep{Leibler2010}.  The variation in mass-to-light ratio between
these age values is about an order of magnitude, as expected from
other galaxy studies.  The 70 Myr and maximum mass-to-light ratios are
listed in Table~\ref{tab:data} for our sample, and in
Table~\ref{tab:other} for the 5 hosts from the literature.

We test the effect of metallicity on the mass-to-light ratio by
considering population synthesis models at $0.02$ and 0.5 Z$_{\odot}$,
which cover the typical range of GRB host galaxy metallicities.  The
resulting variation in mass-to-light ratio is only $\sim 15\%$, with
the lower metallicity models typically yielding systematically smaller 
values (although this effect is redshift-dependent).
This is a much smaller effect than the uncertainty due to the unknown
stellar population age.  In the following, we adopt the mass-to-light
ratios for a metallicity of $0.02$ Z$_{\odot}$.

\begin{figure}[ht]
\centering
\includegraphics[width=1.0\columnwidth]{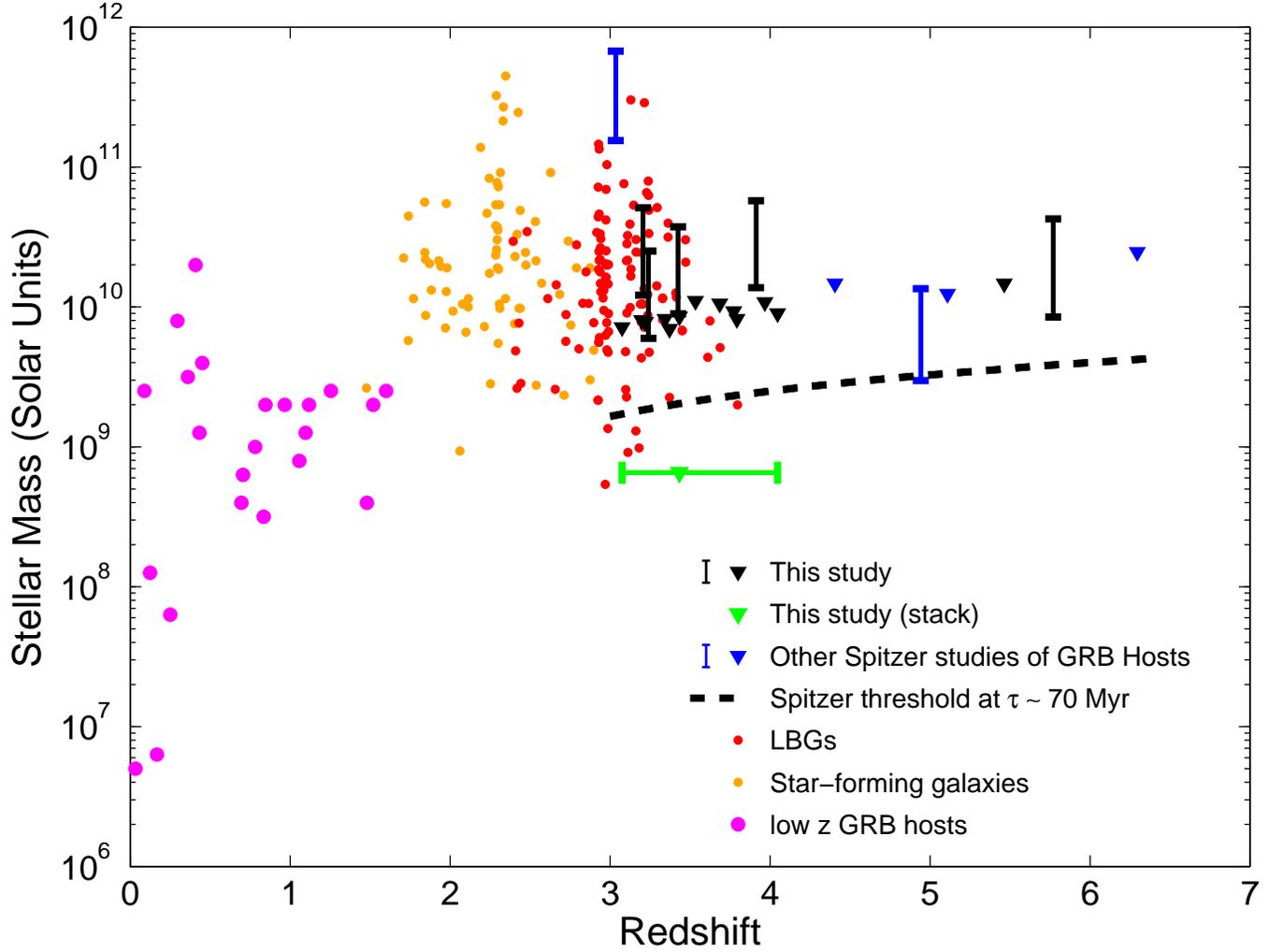}
\caption{Stellar masses plotted as a function of redshift for our
sample (black).  Bars indicate detections with the upper and lower
ends defined by the maximal mass-to-light ratio and a 70 Myr old
stellar population, respectively.  Individual limits (triangles) are
plotted for the maximal masses, while the dashed curve indicates 
the $3\sigma$ limit (0.25 $\mu$Jy) for a 70 Myr old
stellar population.  The stack limit for 11 non-detections at
$z\approx 3.4$ is designated by the green bar (70 Myr old
population).  Also shown are the 5 previous \spitzer\ observations at
$z\gtrsim 3$ from the literature (blue symbols), and low redshift GRB
hosts from the study of \citet{Leibler2010}.  The additional
comparison samples include star forming galaxies at $z\sim 2$ (orange;
\citealt{Shapley2005}) and LBGs at $z\sim 3$ (red;
\citealt{Reddy2006,Maiolino2008,Mannucci2009,Magdis2010}).  The {\it
Spitzer}-detected GRB hosts at $z\sim 3-5$ have similar masses to the
most massive GRB hosts at $z\sim 1$ and to the LBGs at $z\sim 3$.  The
stack limit is similar to the typical masses of GRB hosts at $z\sim
1$.
\label{fig:mass}}
\end{figure}

The inferred masses of our GRB host sample are plotted as a function
of redshift in Figure~\ref{fig:mass} and are listed in
Table~\ref{tab:data}. The maximal masses inferred
for our sample are $(2.5-5.8)\times 10^{10}$ M$_\odot$, while the
typical (maximal) upper limits are $\lesssim 9\times 10^{9}$
M$_\odot$. The masses inferred for a 70 Myr old population
are about $(0.6-1.4)\times 10^{10}$ M$_\odot$, with typical upper
limits of $\lesssim 2\times 10^9$ M$_\odot$. The mass limit from
the stack of 11 GRB hosts is $\lesssim 7\times 10^8$ M$_\odot$ for a
70 Myr old population, and $\lesssim 3\times 10^9$ M$_\odot$ for the
maximal age. We also consider the previous 5 long GRB hosts
at $z\gtrsim 3$ (Table~\ref{tab:other}), including two detections with maximal
masses of $1.4\times 10^{10}$ M$_\odot$ (GRB\,060510B;
\citealt{Chary2007a}) and $6.7\times 10^{11}$ M$_\odot$ (GRB\,080607;
\citealt{Chen2010}) in our $M_*$-$Z$ analysis.  

In comparison to these values, the typical stellar mass of the $z\sim
1$ GRB host sample is about $1.2\times 10^9$ M$_\odot$, similar to
our stack limit.  Similarly, the most massive GRB hosts at
$z\sim 1$ have masses of $\sim 10^{10}$ M$_\odot$, similar to our
detected hosts.  We reach a similar conclusion in comparison to the
LBG sample at $z\sim 3$: the typical stellar masses of the \spitzer\
detected LBGs \citep{Reddy2006,Magdis2010} are about $10^{10}$
M$_\odot$, although some of these galaxies ($\sim 5\%$) have stellar
masses in excess of $10^{11}$ M$_\odot$.  This is similar to the
distribution of our detected hosts and the 5 literature hosts.  Our
stack (maximal) limit falls below the typical stellar mass detection
limit of the LBG sample.  However, in the absence of a detailed mass
function, it is difficult to estimate how the flux limit associated
with the LBG selection compares to our detected fraction.  Regardless
of the exact answer, it is clear that deeper observations of the GRB
host sample will probe lower mass systems than available with the LBG
sample.

\section{The Mass-Metallicity Relation at $z\sim 3-5$}
\label{sec:mz}

\begin{figure}[ht]
\centering
\includegraphics[width=1.0\columnwidth]{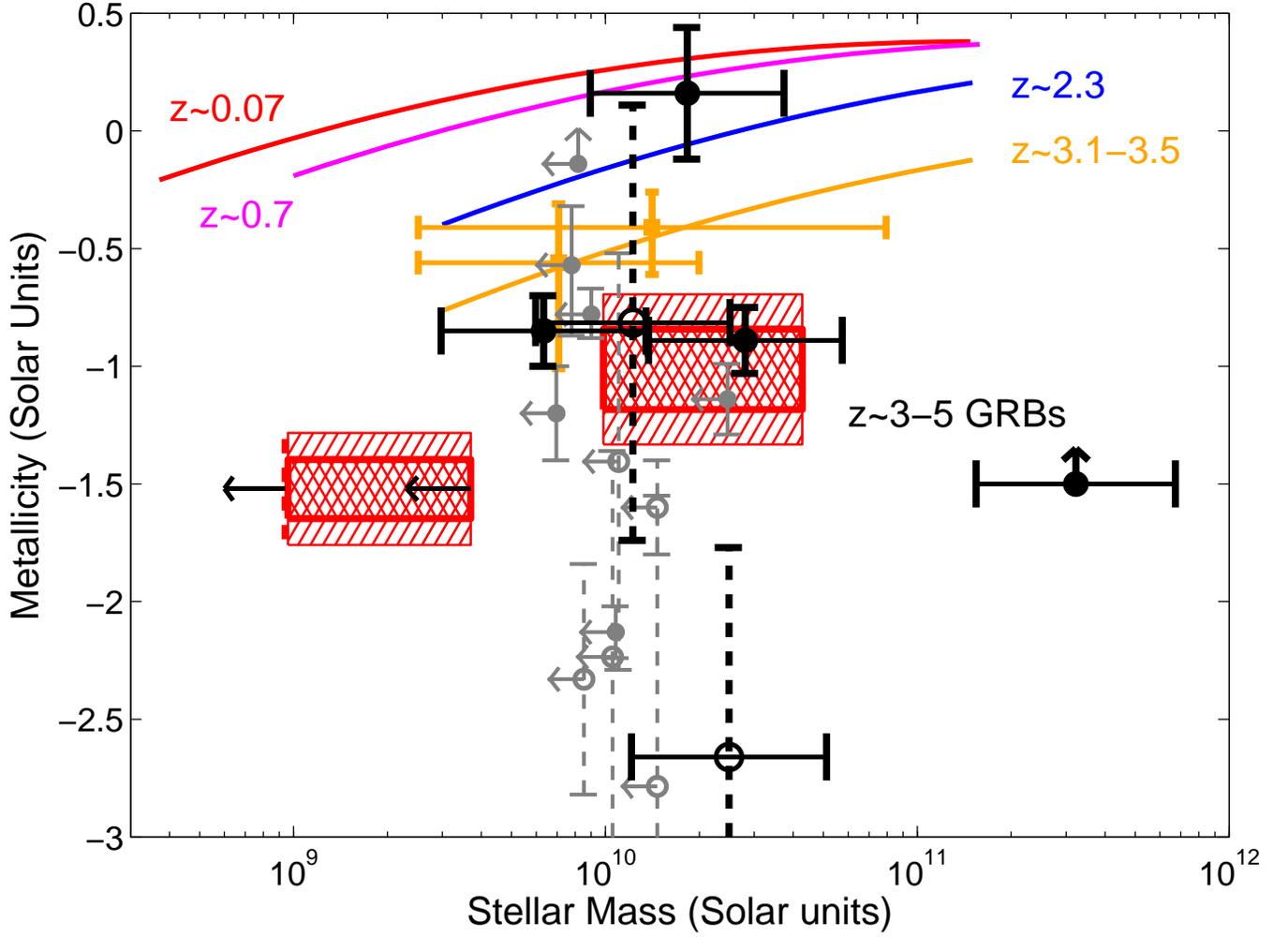}
\caption{Stellar mass plotted as a function of ISM metallicity for our
sample and the 5 previously observed hosts (black=detections;
gray=limits). Confirmed metallicities 
(and one metallicity lower limit) are indicated by filled symbols,
while metallicity ranges are shown by dashed vertical lines 
with open symbols.
The red hatched regions designate $1\sigma$ and $2\sigma$ 
intervals for estimates of the mean metallicity at two mass bins of 
$\sim 2\times 10^{10}$~M$_\odot$ (3.6 $\mu$m detections) and
$\lesssim 3.7\times 10^9$~M$_\odot$ (scaled stack limit - see text).
The red dashed vertical line indicates the upper limit on the mean
mass of the stack for a 70 Myr population.
These data are consistent with a decline in
metallicity with lower stellar mass --- an $M_*$-$Z$ relation.  Also
shown are the relations for $z\sim 0.07$ \citep{Kewley2008}, $z\sim
0.7$ \citep{Savaglio2005}, $z\sim 2.3$ \citep{Erb2006}, and $z\sim
3.1-3.5$ \citep[][filled squares]{Maiolino2008,Mannucci2009}; the relation at $z\lesssim
2.3$ are the re-calibrated values by \citet{Maiolino2008}.  Our two
data regions at $z\sim 3-5$ fall below these relations suggesting that the
$M_*$-$Z$ relation continues to evolve to $z\sim 4$.
\label{fig:MZ}}
\end{figure}

We now turn to the primary investigation of this paper --- the $M_*$-$Z$
relation at $z\sim 3-5$.  In Figure~\ref{fig:MZ} we present the
absorption line metallicities plotted versus the stellar masses
inferred from our \spitzer\ observations (``our sample'').  Also included are the 5 GRB
hosts from previous targeted observations (``literature sample'', Table~\ref{tab:otherZ}).
Of the 18 GRBs in our sample, six have determined [S/H] values, 
five have no metallicity information, and six have upper limits on their
metallicity from non-detections of \ion{S}{2} as well as
lower limits based on \ion{Si}{2} or \ion{Si}{4} detections.
For GRB\,050908, the metallicity upper limit ($Z<10^2$ Z$_{\odot}$)
is not meaningful and we only report a lower limit based on a \ion{C}{2} detection.
For the literature sample, the spectrum of GRB\,080607 exhibits
a saturated \ion{S}{2} line, leading to a lower limit on the metallicity;
GRB\,060223A has both upper and lower limits on the metallicity based on a \ion{S}{2}
non-detection and a \ion{Si}{2} detection, respectively; GRBs
060510B and 050904 have measured metallicities; and GRB\,060522 has no
metallicity information.

Using these values we find a wide range of
metallicities\footnotemark\footnotetext{To transform the sulfur and
oxygen abundances we use the Solar values listed in
\citet{Asplund2005}.} ($Z\approx 0.01-1.5$ Z$_\odot$) for the {\it
Spitzer}-detected GRB hosts, which have stellar masses of $\sim
2\times 10^{10}$ M$_\odot$. This range indicates that at least some
of the hosts have metallicities that are typical of $z\sim 1-2$
galaxies in the same mass range. We note that this range is significantly
larger than the scatter in metallicity observed at low redshift,
which is about 0.4 dex at $\log{M_*}\sim10$ \citep{Tremonti2004}.
Since GRBs probe the metallicities of their host galaxies along random lines of sight
(whereas direct galaxy spectroscopic observations yield luminosity-weighted metallicities),
this larger scatter may be indicative of the intrinsic scatter
in the metallicities of individual star-forming regions in $z\gtrsim 3$ galaxies.
We return to this point in \S\ref{sec:conc}.

To search for an $M_*$-$Z$ relation, 
we divide the GRBs with available metallicity information
(either a metallicity detection or a bounded range) into two mass bins 
--- the 3.6 \mum\ detections with $M_*\sim2\times 10^{10}$ M$_{\odot}$ 
(Group 1: GRBs 060926, 050319, 060707, 060210, 060510B) and the objects included in the stack
(Group 2: GRBs 060526, 030323, 061110B, 060115, 060906, 050730, 060206). 
For each group, we perform a Monte Carlo simulation to estimate the mean metallicity.
We represent metallicity detections by Gaussian random variables 
with a mean equal to the detected metallicity and variance equal to the 
observed uncertainty, and objects with metallicity ranges by uniform distributions.
The simulations yield nearly-Gaussian distributions for the mean metallicity
of objects in both bins, with $\langle Z_{1}\rangle =-1.01 \pm 0.17$ and
$\langle Z_{2}\rangle =-1.52 \pm 0.12$,
where the quoted uncertainties are $1\sigma$ errors on the mean.
The mean metallicities of the two groups are measurably different
at the level of about 1.8$\sigma$.

For Group 1, the mean of the maximum inferred stellar masses is $4.3\times 10^{10}$ M$_{\odot}$,
while that of the masses inferred from the 70 Myr populations is $9.8\times 10^9$ M$_{\odot}$.
To obtain mass estimates for Group 2, we scale our stack limit obtained for 
11 non-detections by $\sqrt{11/7}$. Using the mean maximum mass-to-light ratio
of the objects in Group 2 ($3.7\times10^{10}$ $M_{\odot}/\mu$Jy)
yields an upper limit on the mean stellar mass of
these seven objects of $3.7\times 10^9$ $M_{\odot}$, 
while the mean mass-to-light ratio at 70 Myr ($9.4\times10^{9}$ $M_{\odot}/\mu$Jy)
yields a mass limit of $9.4\times 10^8$ $M_{\odot}$.
The resulting mass ranges together with the corresponding
1$\sigma$ and 2$\sigma$ metallicity ranges for both groups are indicated 
by hatched regions in Figure~\ref{fig:MZ}.
We find that the mean metallicity decreases as a function of stellar mass,
an initial indication of an $M_*$-$Z$ relation.
We note that our averaging of the individual metallicities
at a fixed stellar mass is similar to the approach taken by
\citet{Erb2006} for their $z\sim 2.3$ sample for which they
constructed composite spectra in various mass bins (i.e., they
averaged the spectra, while we average the individual metallicities).

To compare our measurements with the observed $M_*$-$Z$ relations at
lower redshifts, we need to ensure the use of a common stellar IMF and
calibration of the spectral indices used to measure the metallicities.
We use the results of \citet{Maiolino2008} who re-calibrated the
$z\sim 0.07$ relation of \citet{Kewley2008}, the $z\sim 1$ relation of
\citet{Savaglio2005}, and the $z\sim 2.3$ relation of \citet{Erb2006} to
the Salpeter IMF.  In Figure~\ref{fig:MZ} we plot the resulting
$M_*$-$Z$ relations given in \citet{Maiolino2008}, which are of the
form:
\begin{equation}
Z\equiv {\rm [O/H]} = -0.0864 \left(\log{M_*} - \log{M_0}\right)^2 +
K_0 - ({\rm O/H})_{\odot},
\end{equation}
where $M_0$ and $K_0$ are the parameters of the log-parabolic fit to
the re-calibrated data, and $({\rm O/H})_\odot=8.66$
is the solar oxygen abundance \citep{Asplund2005}.  We also include in
Figure~\ref{fig:MZ} the $M_*$-$Z$ relation inferred for LBGs at $z\sim
3.1-3.5$ \citep{Maiolino2008,Mannucci2009}, along with the mean
$M_*$-$Z$ points for $z\sim 3.1$ \citep{Mannucci2009} and $z\sim 3.5$
\citep{Maiolino2008}.  We find that our two points fall below
the observed relations at $z\lesssim 3.5$, providing evidence that the
galaxy $M_*$-$Z$ relation continues to evolve at $z\sim 3-5$, 
with our stack range probing a somewhat lower mass scale than 
the LBG studies at $z\sim 3.1-3.5$.

\section{Discussion and Conclusions}
\label{sec:conc}

We present the first study of the galaxy mass-metallicity relation at
redshifts of $z\sim 3-5$ using GRB afterglow absorption metallicities
and \spitzer\ follow-up observations.  Five of the 20 GRB hosts in our
sample are detected above a 3$\sigma$ flux density threshold of 0.25
$\mu$Jy, corresponding to a typical stellar mass of $\sim 2\times
10^{10}$ M$_\odot$.  We further place a limit of $\lesssim 3\times
10^9$ M$_\odot$ on the non-detected hosts based on a stacking
analysis.

The rest-frame optical luminosities and derived masses are generally
similar to those found for GRB hosts at lower redshifts, but are
larger than for LAEs at similar redshifts.  The comparison to the LBG
population is less certain.  No GRB hosts in our sample are detected
with $M_{\rm AB}\lesssim -23$ mag, while about $20\%$ of the LBG
sample are more luminous than this value.  On the other hand,
integration of the $z\sim 3$ optical luminosity function suggests that
we expect only $\sim 2$ GRB hosts brighter than this limit in our
sample, statistically consistent with zero detections.  The recent
detection of the host galaxy of GRB\,080607 with a luminosity and mass
similar to the most massive LBGs supports the possibility that GRB
hosts and LBGs are not dissimilar \citep{Chen2010}.

We find a wide dispersion in the metallicities of the host galaxies
(inferred mainly from \ion{S}{2}) at a fixed stellar mass of
$\sim 2\times 10^{10}$ M$_\odot$. The mean metallicity at
this mass scale is about $0.1$ Z$_\odot$.  The mean metallicity
associated with 7 of the 11 non-detected hosts, which have an upper limit of
$\lesssim 3.7\times 10^9$ M$_\odot$, is $Z\lesssim 0.03$
Z$_\odot$.  Thus, there appears to be an overall decline in
metallicity with decreasing stellar mass, a hint of an $M_*$-$Z$ relation.
Furthermore, our two points on the $M_*$-$Z$ relation lie below the
relations at lower redshifts, suggesting that the relation continues
to evolve at least to $z \sim 4$.  Clearly, additional observations are
required to confirm and increase the statistical significance of this
result.  A sample of 20 additional GRBs at $z\gtrsim 3$ from 2007
through the present is available for study.  This will allow us to
double the existing sample.

The observed range in metallicities at $M_*\sim2\times10^{10}$ M$_\odot$
appears to be larger than the observed scatter in metallicities 
at similar stellar masses in the nearby universe.
While it is possible that this is a real effect, we caution that
this may be an observational artifact; GRBs probe individual sight lines 
through their host galaxies, whereas traditional methods integrate the spectrum 
over an aperture or slit, thereby averaging over many individual \ion{H}{2} regions 
(weighted by their luminosity). Since we divide galaxies into two groups and 
compute their mean metallicities, in effect we achieve a similar result as 
integrating over the many \ion{H}{2} regions in individual galaxies.

We end the paper with some cautionary notes and future prospects. One
possible source of systematic uncertainty in metallicities as probed
by GRBs lies in the radial abundance profile of the host galaxies.
The Milky Way and M33 display a strong abundance gradient ($\sim
-0.07$ dex kpc$^{-1}$: \citealt{Rolleston2000, Cioni2009}).  Similar
abundance gradients ($\sim -0.05$ dex kpc$^{-1}$) have been found for
\ion{H}{2} regions in nearby spiral galaxies (e.g.,
\citealt{Vila-Costas1992,vanZee1998}).  At higher redshift,
\citet{Jones2010} find a gradient of $-0.3$ dex kpc$^{-1}$ in a lensed
system at $z=2.0$; while this gradient is large in absolute terms,
they clarify that it is similar to gradients in nearby spirals when
the evolution of the effective radius out to $z\sim 2$ in taken into
account.  On the other hand, the LMC and SMC, which may be more
representative of GRB hosts, display almost no radial metallicity
gradient \citep{Cioni2009}.  Since GRBs probe an unknown line of sight
through their hosts, a strong metallicity gradient combined with a
preferred location for the progenitors may lead to a systematic bias
in the resulting metallicity measurements.  However, a distribution of
GRBs that uniformly samples \ion{H}{2} regions within their hosts,
coupled with the potential that low mass galaxies at high redshift
have weak gradients, will negate such a bias.

We note that a similar effect may exist in direct galaxy metallicity
measurements.  This is simply because the measured metallicity is
effectively a luminosity-weighted value, which therefore depends on
the combined radial distribution and luminosities of \ion{H}{2}
regions.  For example, if \ion{H}{2} regions in the outskirts of LBGs
were more luminous, an abundance gradient would lead to a biased
metallicity value.  Thus, assuming that galaxies of similar masses
have similar abundance profiles, we would expect metallicities
determined by GRBs as an ensemble for a given galaxy mass to be
representative of the typical galaxy metallicity at that mass.  We
conclude that the effect of metallicity gradients is minimal when
comparing samples as a function of galaxy mass.

A second concern is the relative calibration of absorption
metallicities (using mainly the sulfur abundance) and nebular emission
line metallicities (using mainly the oxygen abundance).  At present,
the uncertainty in the solar abundance of these two elements
(primarily oxygen) leads to at least $\sim 0.1$ dex uncertainty in the
relative calibration.  Beyond this problem, an additional concern is
that while to first order GRB absorption spectra and nebular lines
both trace regions of star formation, it is unclear how the
luminosity-weighted nebular metallicities relate in detail to the line
of sight GRB metallicities (even in the absence of metallicity
gradients).  Thus, cross-calibration of the metallicities using direct
spectroscopy of GRB hosts is of the utmost importance.  This is
missing at the present.

Looking beyond additional \spitzer\ observations of existing GRB
hosts, which from our work appear to have a detection yield of $\sim
25\%$ at $z\gtrsim 3$, the {\it James Webb Space Telescope} (JWST)
will provide a much deeper view of the $M_*$-$Z$ relation at high
redshift.  For instance, the NIRCam instrument on JWST employing the
3.6 \mum\ wide filter will be able to detect point sources at a
$5\sigma$ flux of about 10\,nJy in a similar integration time to our
existing observations.  This will allow us to detect GRB hosts at
$z\sim 3$ down to a mass of $\sim 10^8$ M$_\odot$, and at $z\sim 6$ to
$\sim 3\times 10^8$ M$_\odot$.  At these limits, we should be able to
detect the bulk of the hosts individually if they are similar to GRB
hosts at $z\sim 1$.  Equally important, the NIRSpec instrument ($1-5$
\mum) will allow us to determine emission-line metallicities for some
of the hosts, and hence to cross-calibrate the afterglow absorption
metallicities.  Thus, with on-going afterglow absorption metallicity
measurements, the GRB sample will continue to play a key role in our
study of high redshift galaxies.

\acknowledgements This work is based on observations made with the
Spitzer Space Telescope, which is operated by the Jet Propulsion
Laboratory, California Institute of Technology under a contract with
NASA. Support for this work was provided by NASA through an award
issued by JPL/Caltech.


\bibliographystyle{apj}
\bibliography{/home/tanmoy/Projects/Edo/Publication/astro-ph/journals_apj,/home/tanmoy/Projects/Edo/Papers/grbhosts}

\clearpage
\begin{deluxetable}{lcccccccc}
\tablecolumns{8}
\tablewidth{0pt}
\tabcolsep0.08in\footnotesize
\tablecaption{{\it Spitzer} Observations and Inferred Properties of
GRB Host Galaxies at $z\sim 3-5$ in this Study \label{tab:data}}
\tablehead{
\colhead{GRB} & 
\colhead{$z$} & 
\colhead{$F_\nu (3.6\,{\rm \mu m})$} & 
\colhead{$P_{\rm cc}$}\tablenotemark{1} &
\colhead{$M_{\rm AB}(780\,{\rm nm})$} & 
\colhead{$(M/F_{\nu})_{70\,{\rm Myr}}$} & 
\colhead{$(M/F_{\nu})_{\rm max}$} & 
\colhead{$M_{\rm 70\,Myr}$} & 
\colhead{$M_{\rm max}$} \\
\colhead{} & 
\colhead{} &       
\colhead{($\mu$Jy)} & 
\colhead{(mag)} & 
\colhead{($10^{10} M_{\odot}/\mu{\rm Jy}$)} & 
\colhead{($10^{10} M_{\odot}/\mu{\rm Jy}$)} & 
\colhead{($10^{10} M_{\odot}$)} &
\colhead{($10^{10} M_{\odot}$)}
}
\startdata
060607  & 3.075                   & $<0.24$        & ---   & $>-20.14$        & 0.69 & 2.97 & $<0.17$        & $<0.71$ \\
020124  & 3.198                   & $<0.26$        & ---   & $>-20.29$        & 0.73 & 3.09 & $<0.19$        & $<0.80$ \\
060926  & 3.206                   & $1.65\pm 0.07$ & 4.3\% & $-22.30\pm 0.07$ & 0.73 & 3.10 & $1.21\pm 0.05$ & $5.12\pm 0.22$  \\
060526  & 3.221                   & $<0.25$        & ---   & $>-20.25$        & 0.74 & 3.12 & $<0.19$        & $<0.78$ \\
050319  & 3.240                   & $0.80\pm 0.09$ & 7.7\% & $-21.53\pm 0.20$ & 0.75 & 3.14 & $0.60\pm 0.07$ & $2.51\pm 0.29$ \\
050908  & 3.344                   & $<0.25$        & ---   & $>-20.31$        & 0.78 & 3.27 & $<0.20$        & $<0.82$ \\
030323  & 3.372                   & $<0.21$        & ---   & $>-20.14$        & 0.79 & 3.32 & $<0.17$        & $<0.70$ \\
060707  & 3.425                   & $1.10\pm 0.10$ & 6.8\% & $-21.96\pm 0.16$ & 0.81 & 3.40 & $0.89\pm 0.08$ & $3.74\pm 0.34$ \\
061110B & 3.433                   & $<0.25$        & ---   & $>-20.35$        & 0.82 & 3.41 & $<0.20$        & $<0.85$ \\
060115  & 3.533                   & $<0.31$        & ---   & $>-20.63$        & 0.85 & 3.56 & $<0.26$        & $<1.10$ \\
060906  & 3.686                   & $<0.28$        & ---   & $>-20.58$        & 0.90 & 3.76 & $<0.25$        & $<1.05$ \\
060605  & 3.773                   & $<0.24$        & ---   & $>-20.45$        & 0.92 & 3.88 & $<0.22$        & $<0.93$ \\
050502  & 3.793                   & $<0.21$        & ---   & $>-20.32$        & 0.93 & 3.90 & $<0.20$        & $<0.82$ \\
060210  & 3.913                   & $1.41\pm 0.10$ & 5.9\% & $-22.42\pm 0.11$ & 0.97 & 4.08 & $1.37\pm 0.10$ & $5.75\pm 0.41$ \\
050730  & 3.968                   & $<0.26$        & ---   & $>-20.62$        & 0.99 & 4.15 & $<0.26$        & $<1.08$ \\
060206  & 4.048                   & $<0.21$        & ---   & $>-20.41$        & 1.02 & 4.29 & $<0.21$        & $<0.90$ \\
060927  & 5.464                   & $<0.21$        & ---   & $>-20.83$        & 1.44 & 6.98 & $<0.30$        & $<1.47$ \\
050814  & 5.77\,\tablenotemark{2} & $0.55\pm 0.06$ & 7.6\% & $-21.94\pm 0.20$ & 1.54 & 7.75 & $0.85\pm 0.09$ & $4.26\pm 0.47$
\enddata
\tablenotetext{1}{Probability of chance coincidence.}
\tablenotetext{2}{This is a photometric redshift \citep{Curran2008}.}
\end{deluxetable}

\clearpage
\begin{deluxetable}{lccccccc}
\tablecolumns{8}
\tablecaption{{\it Spitzer} Observations and Inferred Properties of
GRB Host Galaxies from Previous Studies \label{tab:other}}
\tablewidth{0pt}
\tabcolsep0.08in\footnotesize
\tablehead{
\colhead{GRB} & 
\colhead{$z$} & 
\colhead{$F_\nu (3.6\,{\rm \mu m})$} & 
\colhead{$M_{\rm AB}(780\,{\rm nm})$} & 
\colhead{$(M/F_{\nu})_{70\,{\rm Myr}}$} & 
\colhead{$(M/F_{\nu})_{\rm max}$} & 
\colhead{$M_{\rm 70\,Myr}$} & 
\colhead{$M_{\rm max}$} \\
\colhead{} & 
\colhead{} &       
\colhead{($\mu$Jy)} & 
\colhead{(mag)} & 
\colhead{($10^{10} M_{\odot}/\mu{\rm Jy}$)} & 
\colhead{($10^{10} M_{\odot}/\mu{\rm Jy}$)} & 
\colhead{($10^{10} M_{\odot}$)} &
\colhead{($10^{10} M_{\odot}$)}
}
\startdata
080607\,\tablenotemark{a}  & 3.036 & $22.9\pm 2.1$  &  $-25.07\pm 0.09$ & 0.67 & 2.94 & $15.4\pm 1.4$  & $67.3\pm 6.2$ \\
060223A\,\tablenotemark{b} & 4.406 & $<0.30$        &  $>-20.91$        & 1.13 & 4.89 & $<0.34$        & $<1.47$ \\
060510B\,\tablenotemark{b} & 4.942 & $0.23\pm 0.04$ &  $-20.78\pm 0.17$ & 1.29 & 5.86 & $0.30\pm 0.05$ & $1.35\pm 0.23$ \\
060522\,\tablenotemark{b}  & 5.110 & $<0.20$        &  $>-20.68$        & 1.34 & 6.18 & $<0.27$        & $<1.24$ \\
050904\,\tablenotemark{c}  & 6.295 & $<0.27$        &  $>-21.29$        & 1.69 & 9.12 & $<0.46$        & $<2.46$ 
\enddata
\tablenotetext{a}{\citet{Chen2010}}
\tablenotetext{b}{\citet{Chary2007a}}
\tablenotetext{c}{\citet{Berger2007a}}
\end{deluxetable}

\clearpage
\begin{deluxetable}{lcccccc}
\tablecolumns{7}
\tablecaption{Metallicities from Afterglow Absorption Spectroscopy for
GRB Host Galaxies at $z\sim 3-5$ in this Study \label{tab:Z}}
\tablewidth{0pt}
\tabcolsep0.12in\footnotesize
\tablehead{
\colhead{GRB} & 
\colhead{$z$} & 
\colhead{$\log(N_{\rm HI})$} & 
\colhead{Ion} & 
\colhead{Rest Wavelength} &
\colhead{$\log(N_{\rm Ion})$} & 
\colhead{[Z/H]\tablenotemark{1}} \\
\colhead{} & 
\colhead{} &       
\colhead{(cm$^{-2}$)} & 
\colhead{} &
\colhead{(\AA)} &       
\colhead{(cm$^{-2}$)} & 
\colhead{}        
}
\startdata
060607  & 3.075 & $16.95\pm0.03$\,\tablenotemark{a} &  ---        & ---    & ---                               & ---            \\
020124  & 3.198 & $21.70\pm0.20$\,\tablenotemark{b} &  ---        & ---    & ---                               & ---            \\
060926  & 3.206 & $22.60\pm0.15$\,\tablenotemark{a} &  \ion{S}{2} & 1253.8 & $<15.97$\,\tablenotemark{a}       & $<-1.77$       \\
        &       &                                   &  \ion{Si}{2}& 1526.7 & $>14.56$\,\tablenotemark{a}       & $>-3.55$       \\
060526  & 3.221 & $20.00\pm0.15$\,\tablenotemark{a} &  \ion{S}{2} & various\tablenotemark{2}
                                                                           & $14.58\pm0.25$\,\tablenotemark{c} &$-0.57\pm0.25$\,\tablenotemark{c} \\
050319  & 3.240 & $20.90\pm0.20$\,\tablenotemark{a} &  \ion{S}{2} & 1253.8 & $<16.15$\,\tablenotemark{a}       & $<0.11$        \\
        &       &                                   &  \ion{Si}{4}& 1402.8 & $>14.67$\,\tablenotemark{a}       & $>-1.74$       \\
050908  & 3.344 & $17.60\pm0.10$\,\tablenotemark{a} &  \ion{C}{2} & 1334.5 & $>13.85$\,\tablenotemark{a}       & $>-0.14$       \\
030323  & 3.372 & $21.90\pm0.07$\,\tablenotemark{d} &  \ion{S}{2} & 1253.8 & $15.84\pm0.19$\,\tablenotemark{d} & $-1.20\pm0.20$ \\
060707  & 3.425 & $21.00\pm0.20$\,\tablenotemark{a} &  \ion{S}{2} & 1250.6 & $16.30\pm0.2$\,\tablenotemark{a}  & $0.16\pm0.28$  \\
061110B & 3.433 & $22.35\pm0.10$\,\tablenotemark{a} &  \ion{S}{2} & 1253.8 & $<15.65$\,\tablenotemark{a}       & $<-1.84$       \\
        &       &                                   &  \ion{Si}{2}& 1304.4 & $>15.04$\,\tablenotemark{a}       & $>-2.82$       \\        
060115  & 3.533 & $21.50\pm0.10$\,\tablenotemark{a} &  \ion{S}{2} & 1253.8 & $<16.12$\,\tablenotemark{a}       & $<-0.52$       \\
        &       &                                   &  \ion{Si}{2}& 1526.7 & $>14.72$\,\tablenotemark{a}       & $>-2.29$       \\        
060906  & 3.686 & $21.85\pm0.10$\,\tablenotemark{a} &  \ion{S}{2} & 1253.8 & $<15.63$\,\tablenotemark{a}       & $<-1.36$       \\
        &       &                                   &  \ion{Si}{2}& 1526.7 & $>14.25$\,\tablenotemark{a}       & $>-3.11$       \\
060605  & 3.773 & $18.90\pm0.40$\,\tablenotemark{e} &  ---        & ---    & ---                               & ---            \\
050502  & 3.793 & ---                               &  ---        & ---    & ---                               & ---            \\
060210  & 3.913 & $21.55\pm0.10$\,\tablenotemark{a} &  \ion{S}{2} & 1253.8 & $15.80\pm0.10$\,\tablenotemark{f} & $-0.89\pm0.14$ \\
050730  & 3.968 & $22.10\pm0.10$\,\tablenotemark{a} &  \ion{S}{2} & 1253.8 & $15.11\pm0.04$\,\tablenotemark{g} & $-2.13\pm0.11$ \\
060206  & 4.048 & $20.85\pm0.10$\,\tablenotemark{a} &  \ion{S}{2} & various\tablenotemark{2}                                                          
                                                                  & $15.21\pm0.03$\,\tablenotemark{h} & $-0.78\pm0.1$  \\
060927  & 5.464 & $22.50\pm0.15$\,\tablenotemark{a} &  \ion{S}{2} & 1253.8 & $<16.90$\,\tablenotemark{a}       & $<-1.55$       \\
        &       &                                   &  \ion{Si}{2}& 1260.4 & $>13.99$\,\tablenotemark{a}       & $>-4.02$       \\        
050814  & 5.77\,\tablenotemark{3} & ---             &  ---        & ---    & ---                               & ---           
\enddata
\tablenotetext{1}{Solar abundances are from \citet{Asplund2005}.  The
metallicities have been re-derived from the quoted metal ion column
densities, as necessary.}
\tablenotetext{2}{Derived by simultaneous least-squares fitting of the \ion{S}{2} $\lambda 1250.6$, $\lambda 1253.8$, and $\lambda 1259.5$
transitions.}
\tablenotetext{3}{This is a photometric redshift \citep{Curran2008}.}
\tablenotetext{a}{\citet{Fynbo2009}}
\tablenotetext{b}{\citet{Hjorth2003}}
\tablenotetext{c}{\citet{Thone2010}}
\tablenotetext{d}{\citet{Vreeswijk2004}}
\tablenotetext{e}{\citet{Ferrero2009}}
\tablenotetext{f}{This work}
\tablenotetext{g}{\citet{Ledoux2009}}
\tablenotetext{h}{\citet{Fynbo2006}}
\end{deluxetable}

\clearpage  
\begin{deluxetable}{lcccccc}
\tablecolumns{7}
\tablecaption{Metallicities from Afterglow Absorption Spectroscopy for
GRB Host Galaxies from Previous Studies \label{tab:otherZ}}
\tablewidth{0pt}
\tabcolsep0.12in\footnotesize
\tablehead{
\colhead{GRB} & 
\colhead{$z$} & 
\colhead{$\log(N_{\rm HI})$} & 
\colhead{Ion} & 
\colhead{Rest Wavelength} &
\colhead{$\log(N_{\rm Ion})$} & 
\colhead{[Z/H]\tablenotemark{1}} \\
\colhead{} & 
\colhead{} &       
\colhead{(cm$^{-2}$)} &
\colhead{} &       
\colhead{\AA} &
\colhead{(cm$^{-2}$)} & 
\colhead{}        
}
\startdata
080607  & 3.036 & $22.70\pm0.04$\,\tablenotemark{a} & \ion{S}{2}  & 1250.6         & $>16.3$                          & $>-1.5$ \\
060223A & 4.406 & $21.60\pm0.10$\,\tablenotemark{b} & \ion{S}{2}  & 1253.8 $<15.3$\,\tablenotemark{b}       & $<-1.4$ \\
        &       &                                   & \ion{Si}{2} & 1304.4 & $\sim 15.3$\,\tablenotemark{b}   & $>-1.8$ \\
060510B & 4.942 & $21.30\pm0.10$\,\tablenotemark{b} & \ion{S}{2}  & 1250.6, 1253.8 & $15.6\pm0.1$\,\tablenotemark{b}  & $-0.85\pm0.15$ \\
060522  & 5.110 & $21.00\pm0.30$\,\tablenotemark{b} & ---         & ---            & ---                             & ---            \\
050904  & 6.295 & $\approx 21.6$\,\tablenotemark{c} & \ion{S}{2}  & 1253.8         & $15.6\pm0.15$\,\tablenotemark{d} & $-1.14^{+0.14}_{-0.17}$ 
\enddata
\tablenotetext{1}{Solar abundances are from \citet{Asplund2005}.  The
metallicities have been re-derived from the quoted metal ion column
densities, as necessary.}
\tablenotetext{a}{\citet{Prochaska2009}}
\tablenotetext{b}{\citet{Chary2007a}}
\tablenotetext{c}{\citet{Totani2006}}
\tablenotetext{d}{\citet{Kawai2006}}
\end{deluxetable}

\clearpage
\begin{deluxetable}{ccccccccc}
\tablecolumns{8}
\tablecaption{Field Galaxy Comparison Samples \label{tab:comp}}
\tablewidth{0pt}
\tabcolsep0.08in\footnotesize
\tablehead{  
\colhead{Sample} & 
\colhead{$z$} & 
\colhead{$\langle z\rangle$} & 
\colhead{Detections} & 
\colhead{Limits} &
\colhead{$\langle m_{\rm AB}\rangle$}\tablenotemark{1} &
\colhead{$\langle M_{\rm AB}\rangle$}\tablenotemark{2} & 
\colhead{Depth} \\
\colhead{} & 
\colhead{} &       
\colhead{} & 
\colhead{} &       
\colhead{} &       
\colhead{(mag)} & 
\colhead{(mag)} & 
\colhead{(mag)}
}  
\startdata
\citet{Reddy2006}   & $2.29\le z\le 3.00$            & 2.94 & 32 & 6   & 23.2 & -22.2 & $24.9$ \\
\citet{Reddy2006}   & $3.00\le z\le 3.66$            & 3.22 & 31 & 5   & 23.7 & -21.9 & $24.9$ \\
\citet{Magdis2010}  & $2.34\le z\le 3.00$            & 2.93 & 19 & 3   & 23.4 & -22.0 & $25.5$\\
\citet{Magdis2010}  & $3.00\le z\le 3.45$            & 3.20 & 30 & 6   & 23.6 & -22.0 & $25.5$\\
\citet{Shapley2005} & $1.48\le z\le 2.90$            & 2.30 & 72 & 0   & 22.8 & -22.1 & $R<25.5$\\
\citet{Ono2010}     & $3.1,\,3.7$\,\tablenotemark{a} & 3.40 & 11 & 261 & 23.7 & -22.1 & $24.8$ \\
This work           & $3.0\le z\le 5.8$              & 3.43 & 5  & 13  & 23.8 & -22.0 & $24.8$\,\tablenotemark{b}
\enddata
\tablenotetext{1}{Median apparent magnitudes in the \spitzer\ 3.6
\mum\ band \citep{Reddy2006,Magdis2010,Ono2010} and $K$-band
\citep{Shapley2005}.}
\tablenotetext{2}{Corresponding median absolute magnitudes.}
\tablenotetext{a}{This work on LAEs uses two narrow-band filters tuned
to Ly$\alpha$ at $z\sim 3.1$ and $z\sim 3.7$.}
\tablenotetext{b}{Median apparent magnitude of our 3.6 \mum\
non-detections.}
\end{deluxetable}

\clearpage

\clearpage


\clearpage


\clearpage

\clearpage

\clearpage

\clearpage

\clearpage

\end{document}